\documentclass[showpacs,pra]{revtex4}
\usepackage{amssymb}
\usepackage{graphicx}
\usepackage{epstopdf}


\begin{document}
%\subsection*{}

\title{Optical Lattices as Waveguides and Beam Splitters for Atom Interferometry:  An Analytical Treatment and Proposal of Applications}

\author{Tim Kovachy}
\email{tkovachy@stanford.edu}
\affiliation{Department of Physics, Stanford University, Stanford, California 94305}

\author{Jason M. Hogan}
\email{hogan@stanford.edu}
\affiliation{Department of Physics, Stanford University, Stanford, California 94305}

\author{David M. S. Johnson}
\email{david.m.johnson@stanford.edu}
\affiliation{Department of Physics, Stanford University, Stanford, California 94305}

\author{Mark A. Kasevich}
\email{kasevich@stanford.edu}
\affiliation{Department of Physics, Stanford University, Stanford, California 94305}

\date{\today}

\begin{abstract}
We provide an analytical description of the dynamics of an atom in an optical lattice using the method of perturbative adiabatic expansion.  A precise understanding of the lattice-atom interaction is essential to taking full advantage of the promising applications that optical lattices offer in the field of atom interferometry.  One such application is the implementation of Large Momentum Transfer (LMT) beam splitters that can potentially provide multiple order of magnitude increases in momentum space separations over current technology.  We also propose interferometer geometries where optical lattices are used as waveguides for the atoms throughout the duration of the interferometer sequence.  Such a technique could simultaneously provide a multiple order of magnitude increase in sensitivity and a multiple order of magnitude decrease in interferometer size for many applications as compared to current state-of-the-art atom interferometers.
\end{abstract}

\maketitle

\section{Introduction}

Atom interferometry has opened new frontiers in precision metrology.  Highly sensitive gravimeters, gravity gradiometers, and gyroscopes have been constructed, and promising work has been done to integrate these sensors into a robust apparatus that can operate outside the laboratory with applications in inertial navigation and geodesy \cite{ref:varenna,ref:peters,ref:chung,ref:snadden,ref:mcguirk,ref:gustavson,ref:biedermann,ref:takase,ref:wu}.  Moreover, atom interferometers have been used to make competitive measurements of the fine structure constant \cite{ref:hermann,ref:cadoret}.  Since atom interferometric measurements of the fine structure constant do not assume the validity of Quantum Electrodynamics (QED), while determinations of the fine structure constant based on measurements of the electron magnetic moment do make this assumption, comparison between the results of these two methods provides a stringent test of QED \cite{ref:hanneke,ref:gabrielse}.  In addition, an experiment to test Einstein's Equivalence Principle with unprecedented precision is underway \cite{ref:gr}, and atom interferometric gravitational wave detectors offer the possibility to study gravitational radiation in frequency ranges complementary to LIGO and LISA \cite{ref:gravitywaves}.  

Atom interferometers have traditionally relied on matter gratings or light pulses to act as beam splitters and mirrors for matter waves, with atomic wave packets traveling freely between these interaction zones.  Light-pulse schemes using either Raman pulses (where the internal state of the atom is changed) or Bragg pulses (where the internal state of the atom remains unchanged) have been implemented, such as those described in \cite{ref:kasevich,ref:multiphoton,ref:giltner,ref:torii}.  For a number of applications of light-pulse atom interferometers,  such as measurements of gravity and rotation, the sensitivity is proportional to the separation in momentum that can be attained between the two arms \cite{ref:berman}.  In measurements of the fine structure constant, the sensitivity scales as the square of this separation \cite{ref:hermann}.  Therefore, significant efforts have been devoted to the development of Large Momentum Transfer (LMT) beam splitters.  LMT beam splitters achieving momentum splittings of $24 \hbar k$ using multi-photon Bragg pulses have recently been demonstrated \cite{ref:multiphoton}.  However, the required laser intensities to make significant improvements on this result may prove to be prohibitive \cite{ref:braggtheory}.  In contrast, LMT beam splitters that use several two-photon Bragg pulses or a multi-photon Bragg pulse of relatively small order to separate the two arms of the interferometer in momentum space, followed by the acceleration of one of the arms with an optical lattice, could potentially provide multiple order of magnitude increases in attainable momentum separations with relatively modest laser intensity requirements.  

An atom interferometer that uses this method has been successfully operated in a proof of principle experiment (with a maximum demonstrated momentum splitting of $12 \hbar k$) \cite{ref:denschlag}.  In a separate experiment, an atom interferometer with  $10 \hbar k$ LMT beam splitters has been realized using a similar technique \cite{ref:clade}.  Alternatively, both arms of the interferometer could be simultaneously accelerated in opposite directions by two different optical lattices after the initial splitting.  Using this second scheme, an interferometer with $24 \hbar k$ LMT beam splitters that achieves 15$\%$ contrast and an individual beam splitter that provides an $88 \hbar k$ momentum separation have been demonstrated \cite{ref:muellerlattice}.

The utility of atom interferometry hinges upon the ability to precisely calculate the phase accumulated along the different arms of an interferometer \cite{ref:bongs,ref:dubetsky,ref:borde}, of which the phase acquired during interactions of the atoms with light is an important component.  Indeed, the phase obtained by an atom during a Raman or Bragg pulse is well-understood \cite{ref:varenna,ref:braggtheory,ref:moler}.  Analogously, in order to take full advantage of the potential of lattice beam splitters, we must have a detailed understanding of the phase evolution of an atom in an optical lattice.  In this paper, we provide a rigorous analytical treatment of this problem.  To our knowledge, such a treatment has not been previously presented in the literature.

\begin{figure}
\includegraphics[width=5.00in]{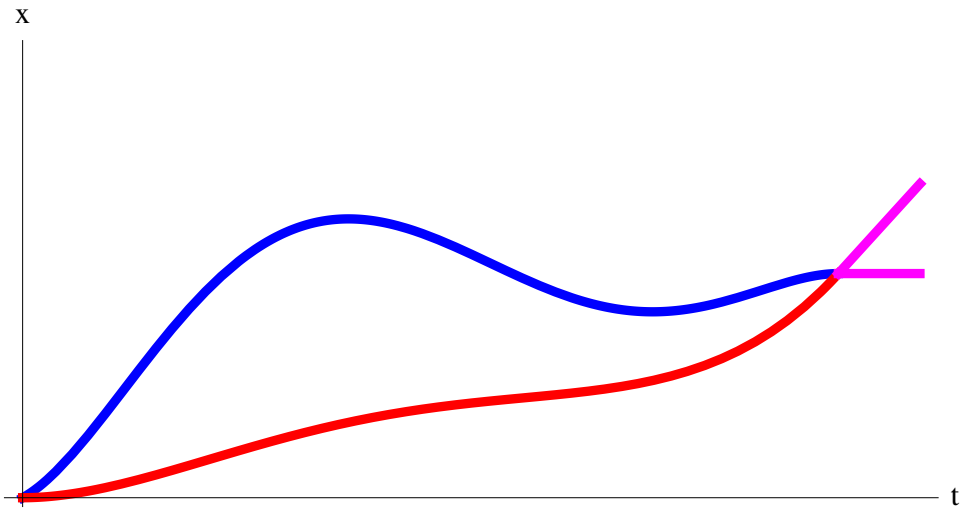} 
\caption{(color online) Schematic of a guided lattice interferometer.  The trajectories of the two arms are guided by two separate optical lattices for the duration of the interferometer sequence, resulting in a fully confined interferometer.  The two paths shown in the figure are the trajectories of the two portions of the atomic wavepacket, which respectively correspond to the two arms of the interferometer.} \label{fig:guidedinterf}
\end{figure}

Based on this analysis, we propose atom interferometer geometries in which optical lattices are used to continuously guide the atoms, so that the atomic trajectories are precisely controlled for the duration of the interferometer sequence, with a different lattice guiding each arm of the interferometer (as illustrated in Fig. \ref{fig:guidedinterf}).  We point out here a distinction in terminology between a lattice waveguide and a lattice beam splitter.  Here, a lattice waveguide is the use of a lattice to continuously control the trajectory of an arm of an atom interferometer.  We note that two separate lattice waveguides can independently control the two arms of an interferometer, or a single lattice waveguide can simultaneously control both arms.  In contrast, a lattice beam splitter is an interaction of relatively short time (in comparison to a waveguide) with the primary purpose of splitting the arms of the interferometer in momentum space rather than providing continuous trajectory control.  The underlying physics behind lattice waveguides and lattice beam splitters is the same, and they can be treated with a common formalism.  A single lattice waveguide that simultaneously transfers $1600 \hbar k$ of momentum to the two arms of a Ramsey-Bord\'{e} interferometer has been previously achieved in \cite{ref:cadoret}.  However, to our knowledge, our idea of using optical lattice waveguides to create a fully confined atom interferometer has not been previously considered.  

Our analysis indicates that these lattice interferometers will offer unprecedented sensitivities for a wide variety of applications and that they will be able to operate effectively over distance scales previously considered too small to be studied by precision atom interferometry.  For example, one particularly interesting configuration involves using two optical lattice waveguides to continuously pull the two arms of the interferometer apart, subsequently holding the two arms a fixed distance from each other in a single lattice waveguide that is common to the two arms, and then using two lattice waveguides to recombine the arms.  Such a configuration could be used, for instance, as a gravimeter.  The sensitivity of lattice interferometers is illustrated by the fact that, given the experimental parameters stated in \cite{ref:varenna} ($10^7$ atoms/shot and $10^{-1}$ shots/s), a shot noise limited lattice gravimeter whose arms are separated by 1 m for an interrogation time of 10 s has a sensitivity of $10^{-14}$ $g$/Hz$^{1/2}$.  We perform phase shift calculations for these lattice interferometers using the theoretical groundwork formulated in this paper, and we discuss how lattice interferometers can both exceed the performance of conventional atom interferometers in many standard applications and expand the types of measurements that can effectively be carried out using atom interferometry.

The paper is organized as follows.  Sec. II. describes the Hamiltonian for an atom in an optical lattice in the different frames we use in the paper.  Sec. III. discusses the phase evolution of an atom in an optical lattice under the adiabatic approximation.  Sec. IV. introduces the formalism of perturbative adiabatic expansion to calculate corrections to the adiabatic approximation, and Sec. V. applies this formalism to calculate phase corrections to a lattice beam splitter.  Sec. VI. proposes a number of interferometer geometries that make use of lattice manipulations of the atoms.  The main results of the paper are Eqs. (\ref{eqn:phasecorrection}) and (\ref{eqn:phasecorrectionref}), which show how to obtain analytical corrections to the lowest order phase shift estimates.  These corrections are surprisingly large, and understanding them is vital to realizing the full accuracy of the sensor geometries proposed in Sec. VI., as well as other geometries utilizing optical lattice manipulations of the atoms.  For example, the gravitational wave detector proposed in \cite{ref:gravitywaves} will likely make use of lattice beam splitters and/or waveguides.  Previously, the phase evolution induced by lattice manipulations was not sufficiently well-understood to allow for a detailed design of the atom optics system or an estimation of the corresponding systematic effects.

\section{The Hamiltonian in Different Frames}

An optical lattice is a periodic potential formed by the superposition of two counter-propagating laser beams.  Atoms can be loaded into the ground state of the lattice by ramping up the lattice depth adiabatically, and the lattice can then be used to impart momentum to the atoms and/or to control the atoms' trajectories.  Optical lattices are thus a useful tool for atom optics.

\begin{figure}
\includegraphics[width=7.00in]{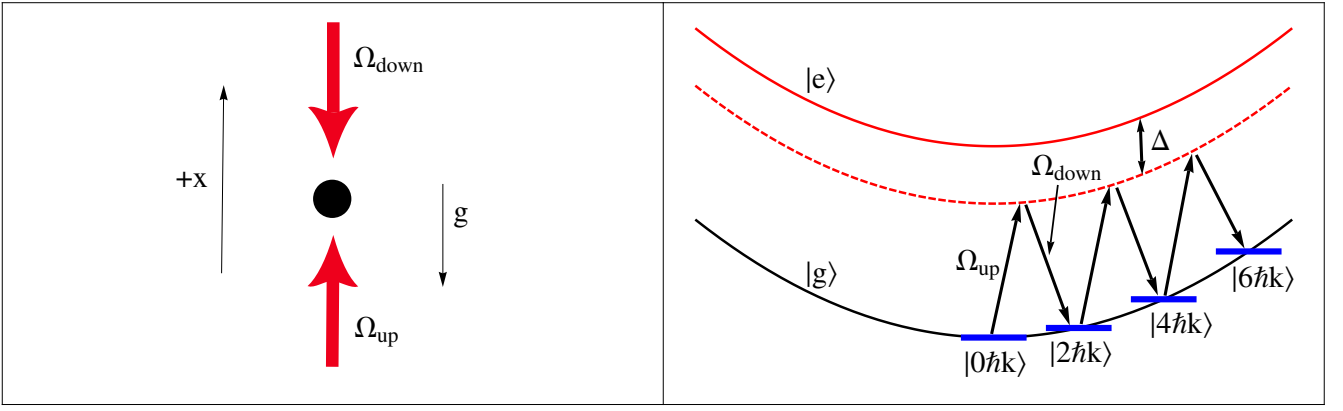} 
\caption{(color online) The physical setup for applying a periodic potential.  We expose atoms to counter-propagating laser beams with respective single-photon Rabi frequencies $\Omega_{\text{up}}$ and $\Omega_{\text{down}}$.  The lasers are detuned from the transition between the atom's internal ground state and excited state so that the atom's external momentum states are coupled through two-photon transitions, creating an effective lattice potential.} \label{fig:setup}
\end{figure}

We begin our discussion of the lattice-atom interaction by finding a useful form for the Hamiltonian.  As is typical for many applications of atom interferometry, to minimize decoherence we assume that we work with atomic gases dilute enough so that the effects of atom-atom interactions are negligible.  We first consider the Hamiltonian in the lab frame, where for now we assume a vertical configuration with constant gravitational acceleration $g$ so that we have a gravitational potential given by $mgx$.  We expose the atom to a superposition of an upward propagating beam with phase $\phi_{\text{up}}(t)$ and a downward propagating beam with phase $\phi_{\text{down}}(t)$, which couples an internal ground state $\left | g \right>$ to an internal excited state $\left | e \right>$.  The two-photon Rabi frequency is $\Omega(t) \equiv \frac{\Omega_{\text{up}}(t) \Omega_{\text{down}}(t)}{2 \Delta}$, where we let $\Omega_{\text{up}}(t)$ denote the single-photon Rabi frequency of the upward propagating beam, $\Omega_{\text{down}}(t)$ denote the single-photon Rabi frequency of the downward propagating beam, and $\Delta$ denote the detuning from the excited state.  We depict the physical setup in Fig. \ref{fig:setup}.  Making the rotating wave approximation and adiabatically eliminating the excited state as is standard procedure \cite{ref:berman}, we obtain the following Hamiltonian--where the periodic term in the potential arises from a spatially varying AC stark shift and where $k$ is the magnitude of the wave vector of the laser beams \cite{ref:wicht,ref:peik}:

\begin{equation}\label{xx}
\ \hat{H}_{\text{Lab}} =\frac{\hat{p}^2}{2m} + 2 \hbar \Omega(t) \sin^2 \left[k\hat{x}-\frac{1}{2}(\phi_{\text{up}}(t) - \phi_{\text{down}}(t))\right]+mg\hat{x}
\end{equation}

\noindent Note that where the difference between the frequency of the upward propagating beam and the frequency of the downward propagating beam is denoted by $\Delta \omega(t)$, we will have the relation $\Delta \phi(t) \equiv \phi_{\text{up}}(t) - \phi_{\text{down}}(t) = \int_{0}^{t}\Delta \omega(t^{ \prime})d t^{\prime}+ \phi_{\text{up}}(0) - \phi_{\text{down}}(0)$.  For a given $\Delta \phi(t)$, the lattice standing wave will be translated by $D_{\text{Lab}}(t) \equiv \frac{\Delta \phi(t)}{2k}$ in the $x$ direction from the origin.  Thus, the velocity of the lattice in the lab frame is:

\begin{equation}
v_{\text{Lab}}(t) = \frac{d}{dt} D_{\text{Lab}}(t) = \frac{\Delta \omega(t)}{2k}
\end{equation}

\noindent and we rewrite the lab frame Hamiltonian as:

\begin{equation}
\ \hat{H}_{\text{Lab}} =\frac{\hat{p}^2}{2m} + 2 \hbar \Omega(t) \sin^2 \left[k\hat{x}-k D_{\text{Lab}}(t) \right]+mg\hat{x}
\end{equation}

\noindent  In order to most readily describe the dynamics of an atom in an accelerating optical lattice, it is useful to work in momentum space.  The $mg\hat{x}$ term that appears in the lab frame Hamiltonian makes such an approach difficult, especially when considering non-adiabatic corrections to the phase shift.  However, we can change frames by performing a unitary transformation in order to obtain a Hamiltonian that is easier to handle analytically.  In the end, we will see that approaching the problem from the point of view of dressed states provides a convenient Hamiltonian for our purposes.  We consider the transformation procedure from the lab frame to the dressed state frame in Appendix A, where we also introduce an intermediate frame that freely falls with gravity (which we call the freely falling frame).  We note that the general form of the unitary transformations considered in Appendix A as well as the specific transformations to the different frames we consider can also be found in the Appendix of \cite{ref:peik}.

It is convenient to absorb the initial velocity $v_0$ of the atom in the lab frame into the dressed state frame, so that velocity $v_0$ in the lab frame corresponds to velocity zero in the dressed state frame.  The Hamiltonian in the dressed state frame is, as derived in Appendix A:

\begin{equation}
\ \hat{H}_{\text{DS}} =\frac{\hat{p}^2}{2m} - (v_{\text{Lab}}(t) + gt - v_0) \hat{p} + 2 \hbar \Omega(t) \sin^2 \left(k\hat{x} \right)
\end{equation}

Now, we will show how working in momentum space allows us to represent $\hat{H}_{\text{DS}}$ as an infinite dimensional, discrete matrix.  This matrix is discrete because the optical lattice potential term, $V_0 \sin^2 \left(k\hat{x} \right)$, only couples a momentum eigenstate $\left | p \right >$ to the eigenstates $\left | p + 2\hbar k \right >$ and $\left | p -2 \hbar k \right >$ \cite{ref:peik}.  For the moment, we will examine the evolution of individual eigenstates of the dressed state Hamiltonian $\hat{H}_{\text{DS}}$.  These eigenstates reduce to single momentum eigenstates $\left | p \right >$ when $\Omega =  0$.  The knowledge of how each of these eigenstates evolves under $\hat{H}_{\text{DS}}$ will allow us to describe the dynamics of an entire wavepacket.  For the moment, we will only consider momentum eigenstates corresponding to an integer multiple of $2 \hbar k$, since we have boosted away the initial velocity $v_0$ of the atom in the lab frame.  We note that it is always possible to transform to a particular dressed state frame in which a given momentum eigenstate in the lab frame corresponds to zero momentum in that dressed state frame.  The results we derive here can thus be readily generalized to arbitrary momentum eigenstates in a wavepacket, as we discuss in greater detail in Appendix B.

We consider a discrete Hilbert space spanned by the momentum eigenstates $\left | 2n\hbar k \right >$ for integers $n$, so that we can express any vector in this Hilbert space as:

\begin{equation}
 \left | \Psi(t) \right > = \sum_{n=- \infty}^{\infty} c_n(t) \left | 2n\hbar k \right >
 \end{equation}
 
\noindent  Since this Hilbert space is discrete, it is natural to adopt the normalization convention that $\left < 2m\hbar k | 2n\hbar k \right > = \delta_{mn}$.  When considered as an operator acting on this discrete Hilbert space, $\hat{H}_{\text{DS}}$ can be written as \cite{ref:braggtheory,ref:malinovsky}:

\begin{eqnarray}
\nonumber \hat{H}_{\text{DS}}^{\text{discrete}} &=&  \sum_{n=- \infty}^{\infty} \frac{(2n\hbar k)^2}{2 m} \left | 2n\hbar k \right > \left < 2n\hbar k \right | - \sum_{n=- \infty}^{\infty} (v_{\text{Lab}}(t) + gt - v_0) (2n \hbar k) \left | 2n\hbar k \right > \left < 2n\hbar k \right | \\
\nonumber &-& \sum_{n=- \infty}^{\infty} \hbar \frac{\Omega(t)}{2} \left ( \left | 2n\hbar k \right > \left < 2(n-1)\hbar k \right | +\left | 2n\hbar k \right > \left < 2(n+1)\hbar k \right | \right ) \\
\end{eqnarray}

\noindent where we drop the common light shift.  Now, it is convenient to introduce the recoil frequency $\omega_{r} \equiv \frac{E_r}{\hbar} = \frac{\hbar k^2}{2m}$ and the recoil velocity $v_r \equiv \frac{\hbar k}{m}$.  In order to make our notation as compact as possible, we will be interested in the quantity $\alpha(t) \equiv \frac{v_{\text{Lab}}(t) + gt - v_0}{v_r}$,  which is the velocity of the lattice in the dressed state frame in units of $v_r$.  We can express the second term of $\hat{H}_{\text{DS}}^{\text{discrete}}$ in a useful way by noting that $(v_{\text{Lab}}(t) + gt - v_0) 2n \hbar k = \frac{v_{\text{Lab}}(t) + gt - v_0}{v_r} v_r 2n \hbar k =4 n \alpha(t) E_r$.  Furthermore, we define $\tilde{\Omega}(t) \equiv \frac{\Omega(t)}{8 \omega_r}$.  We can now write the discrete Hamiltonian in a simplified form:

\begin{equation}
\nonumber \hat{H}_{\text{DS}}^{\text{discrete}} =4 E_r \sum_{n=- \infty}^{\infty} [n^2 \left | 2n\hbar k \right > \left < 2n\hbar k \right | -  n \alpha(t)  \left | 2n\hbar k \right > \left < 2n\hbar k \right | - \tilde{\Omega}(t) \left ( \left | 2n\hbar k \right > \left < 2(n-1)\hbar k \right | +\left | 2n\hbar k \right > \left < 2(n+1)\hbar k \right | \right )] \\
\end{equation}

\noindent The matrix elements of this Hamiltonian are:

\begin{equation} \label{eqn:hamiltonianelements}
H_{mn} \equiv \left < 2m\hbar k \right | \hat{H}_{\text{DS}}^{\text{discrete}} \left | 2n\hbar k \right > = 4 E_r \left [ \left (n^2 - n \alpha(t) \right ) \delta_{mn} - \tilde{\Omega}(t) \left(\delta_{m,n+1}+ \delta_{m,n-1} \right ) \right ]
\end{equation}

\noindent In matrix notation, the Schrodinger equation for the discrete Hilbert space takes the form 
$i \hbar \frac{\partial}{\partial t} \vec{\Psi}(t) = H \vec{\Psi}(t)$, where we let $H$ be the matrix whose element in the $m$th row and $n$th column is given by $H_{mn}$ and we let $\vec{\Psi}(t)$ be the column vector whose $n$th entry is $c_n(t)$.

\section{Phase Evolution Under the Adiabatic Approximation}

\indent Now that we have determined the Hamiltonian matrix $H$, we have the appropriate machinery in place to describe the phase evolution of an atom in an optical lattice.  We consider the process in which momentum is transferred to the atom through Bloch oscillations.  Reference \cite{ref:peik} provides a thorough and insightful description of Bloch oscillations in a number of different pictures.  Given our choice of Hamiltonian, we work in the dressed state picture, which is discussed in Sec. IV.B. of \cite{ref:peik}, making extensive use of Bloch's theorem, the concept of Brillouin zones,  and the band structure of the lattice \cite{ref:denschlag}.  As in the previous discussion, we consider the evolution of single eigenstates of the dressed state Hamiltonian, noting that we can easily generalize our results to the case of a wave packet of finite width, as we address in Appendix B.

Initially, we consider the system to be in a momentum eigenstate.  First, we adiabatically ramp up the lattice depth by increasing the laser power so that we load the system into an eigenstate of the Hamiltonian.  For the purposes considered here, we want the system to enter the ground eigenstate (corresponding to the zeroth band of the lattice).  In order for this to be achieved, a resonance condition must be met, which states that the velocity of the lattice must match the velocity of the atom to within $v_r$.   The loading will be adiabatic if the adiabatic condition $\left | \left < 1 \right | \dot{H} \left | 0 \right > \right | \ll \frac{(\varepsilon_1-\varepsilon_0)^2}{\hbar}$ is satisfied, where $\left | 0 \right >$ and $\left | 1 \right >$ respectively denote the ground state and the first excited state of the Hamiltonian \cite{ref:peik,ref:denschlag}.  This condition will be easier to meet near the center of the band (where the velocity of the lattice is identical to the velocity of the atom), because the energy gap $\varepsilon_1-\varepsilon_0$ between the zeroth band (which corresponds to ground state) and the first band (which corresponds to the first excited state) becomes smaller as the velocity difference between the lattice and the atom becomes larger (which corresponds to moving toward the border of the first Brillouin zone).  The resonance condition is discussed further in Appendix C.

In the lab frame, the atom accelerates under gravity, increasing the deviation between its velocity and the lattice velocity during the loading process.  In the freely falling and dressed state frames, in which gravity is boosted away, this corresponds to the lattice accelerating upward while the atom remains at rest.  This effect can negatively impact the loading efficiency if the loading sequence is sufficiently long so that the accrued velocity difference becomes a significant fraction of $v_r$.  In such a scenario, the effect can be ameliorated by accelerating the lattice in the lab frame to fall with the atom, which corresponds to the lattice velocity remaining constant in the freely falling and dressed state frames.  Furthermore, we note that in the case of a lattice LMT beam splitter, the lattice should only be resonant with one arm of the interferometer, so that negligible population from the other arm is affected.  Otherwise, the signal could be distorted by multi-path interference, causing a systematic error in the estimation of the interferometer phase shift.  Conditions for when the negative effects of off-resonant lattices can be avoided can be estimated using the Hamiltonian matrix for an off-resonant lattice given in Eq. (\ref{eqn:offresmatrix}).  We discuss off-resonant lattices quantitatively and in more detail in Sec. VI.

After the adiabatic loading of the atom into the ground state of the Hamiltonian, the frequency difference between the laser beams is swept to accelerate the lattice, periodically imparting momentum to the atom in units of $2 \hbar k$.  In the dressed state frame, this phenomenon can be understood in terms of avoided line crossings, which occur because the coupling between the atom and the laser beams lifts the degeneracy at the crossing points.  We refer the reader to Fig. 10 of \cite{ref:peik} for a clear illustration of these avoided crossings.  As the frequency difference is swept so that the system passes through the avoided crossings, the system remains in the ground state of the dressed state Hamiltonian as long as the process is adiabatic.  Consequently, at each of the avoided crossings, the momentum of the atom increases by $2 \hbar k$, which corresponds to a Bloch oscillation.  Finally, after the acceleration of the lattice, the frequency difference is held constant while the lattice depth is adiabatically ramped down, delivering the system into the momentum eigenstate $\left | p_i + 2N\hbar k \right >$, where $p_i$ is the momentum before the Bloch oscillations and $N$ is the number of Bloch oscillations. 

\begin{figure} \includegraphics[width=7.5in]{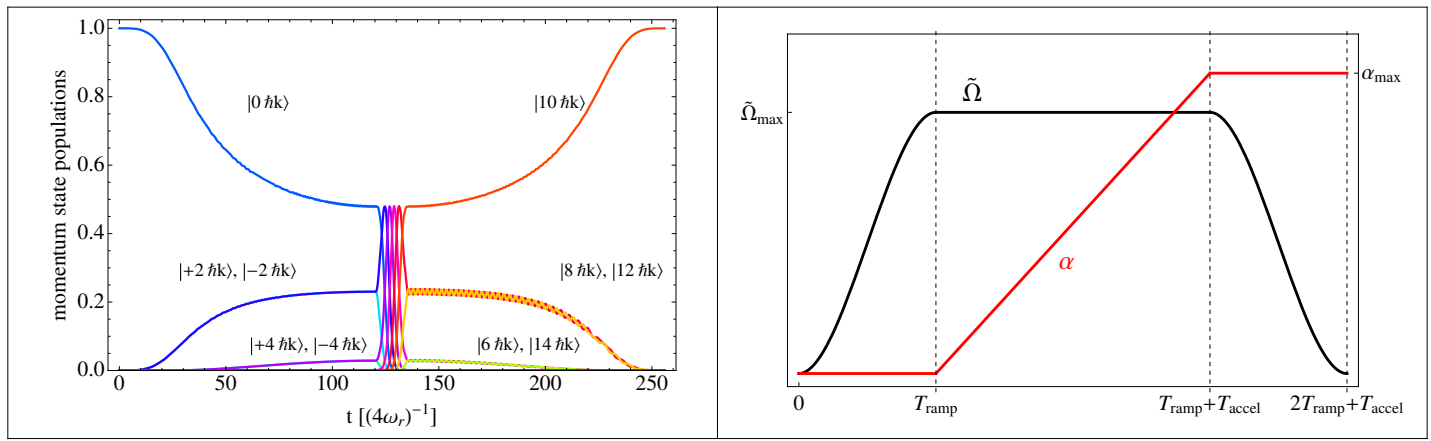} 
\caption{(color online) The left panel shows a numerical simulation of a lattice acceleration in momentum space that transfers $10 \hbar k$ of momentum.  The lattice depth and velocity are shown as a function of time in the right panel, where in this particular case the relevant parameters are $T_{\text{ramp}} = 120 (4 \omega_r)^{-1}$, $T_{\text{accel}} = 16 (4 \omega_r)^{-1}$, $\tilde{\Omega}_{\text{max}} = \frac{19.5}{8}$,  and $\alpha_{\text{max}} = 10$ (corresponding to a final lattice velocity of 10 $v_r$).  First, we adiabatically ramp up the lattice to a depth of $19.5 E_{r}$ so that the lattice is loaded into the ground state of the dressed state Hamiltonian.  Subsequently, we accelerate and ramp down the lattice, leaving the atom in a single momentum eigenstate.}
\label{fig:latticeaccel}
\end{figure}

Fig. \ref{fig:latticeaccel} depicts the lattice depth and velocity as functions of time for the process described above and shows a numerical simulation of a particular instance of this process:  the adiabatic loading of the lattice from an initial state $\left | 0 \hbar k \right >$, the transfer of $10 \hbar k$ of momentum through 5 Bloch oscillations, and the ramping down of the lattice to deliver the system into the final state $\left | 10 \hbar k \right >$.

Since under the adiabatic approximation we assume that the atom always stays in the ground state of the Hamiltonian, the phase $\phi(t)$ of the atom evolves as follows:

\begin{equation} \label{eqn:adiabaticphase}
\ \phi(t)-\phi_0 =  - \frac{1}{\hbar} \int_{t_0}^{t} \varepsilon_0(t^{\prime}) d t^{\prime}
\end{equation}

\noindent where $\varepsilon_0(t)$ is the instantaneous ground state eigenvalue of the Hamiltonian and $\phi_0$ is the initial phase of the atom.  In addition to Eq. (\ref{eqn:adiabaticphase}), there is also a Berry's phase term \cite{ref:griffiths}.  However, this term is zero for a linear external potential.  Therefore, there is no contribution from the Berry's phase under the semiclassical approximation, as long as the external potential is treated as linear--we discuss the validity and ramifications of this approximation at the end of Appendix A.  We note that any such contribution would arise from the residual external potential terms of the dressed state Hamiltonian that are non-linear in $x$ (we collectively denote these terms as $V^{\prime}$ in Appendix A and explain why they can often be neglected).

We now consider how the eigenvectors and eigenvalues of $H$ change with $\alpha$, which we recall is the dimensionless velocity of the atom in the dressed state frame.  Say that the eigenvalues of $H(\alpha,\tilde{\Omega})$ are given by $\varepsilon_n(\alpha,\tilde{\Omega})$ with corresponding eigenvectors $\vec{\Psi}_n(\alpha,\tilde{\Omega})$, where the index $n$ runs from $0$ to $\infty$.  We choose to index the eigenvalues so that $\varepsilon_n(\alpha,\tilde{\Omega})$ denotes the the $n$th eigenvalue labeled in order of increasing value.  Moreover, we let $c^{(n)}_j(\alpha,\tilde{\Omega})$ be the $j$th element of the column vector $\vec{\Psi}_n(\alpha,\tilde{\Omega})$, so that  $\left | \Psi_n(\alpha,\tilde{\Omega}) \right > = \sum_{j=- \infty}^{\infty} c^{(n)}_j(\alpha,\tilde{\Omega}) \left | 2j\hbar k \right >$.  The transformation properties of the eigenvectors and eigenvalues under changes in $\alpha$ can be deduced from Bloch's theorem.  It can be shown that when the lattice velocity is increased by $2v_r = \frac{2 \hbar k}{m}$, which corresponds to $\alpha$ being increased by two, while $\tilde{\Omega}$ is kept fixed, the new eigenvectors can be obtained through the following relation \cite{ref:kovachy}:

\begin{equation}
\ c^{(n)}_j(\alpha + 2,\tilde{\Omega}) =  c^{(n)}_{j-1}(\alpha,\tilde{\Omega})
\end{equation}

\noindent  This simply represents a shift of the wavefunction in momentum space by $2 \hbar k$, which is exactly what we expect, since increasing the lattice velocity by $2v_r$ corresponds to undergoing a single Bloch oscillation. The dependence of the eigenvalues on the lattice velocity can be expressed as follows:

\begin{equation} \label{eqn:evalue}
\varepsilon_n(\alpha,\tilde{\Omega}) = - E_r \alpha^2 +  \varepsilon_n(0,\tilde{\Omega}) + p_n(\alpha,\tilde{\Omega})
\end{equation}

\noindent  where $p_n(\alpha,\tilde{\Omega})$ is periodic in $\alpha$ such that  $p_n(\alpha+2 m,\tilde{\Omega})= p_n(\alpha,\tilde{\Omega})$ for integer $m$ holds for all $\alpha$ and $p_n(\alpha,\tilde{\Omega})$ vanishes when the condition $\alpha = 2m$ holds.  This periodicity follows directly from Bloch's theorem, since increasing $\alpha$ by $2m$ corresponds to increasing the lattice velocity by $2mv_r$, meaning that $\alpha$ and $\alpha + 2m$ are at the same point in the first Brillouin zone.  Note that the dependence of the eigenvalue on $\tilde{\Omega}$ can be calculated using the truncated matrix approximation discussed in Sec. IV.  The relevance of this dependence to the phase shift of an interferometer and how this dependence varies with momentum are discussed in Sec. V. and Appendix B.   

The first term in Eq. (\ref{eqn:evalue}) has a simple physical interpretation.  Where $v_{\text{Lattice}}(t)$ is the velocity of the lattice in the dressed state frame (so that $v_{\text{Lattice}}(t) = \alpha v_r$), note that:

\begin{equation}
E_r \alpha^2=\frac{\hbar^2 k^2}{2m}\left(\frac{v_{\text{Lattice}}(t)}{\frac{\hbar k}{m}} \right)^2 = \frac{1}{2}mv_{\text{Lattice}}(t)^2
\end{equation}

\noindent which is simply the kinetic energy of an atom traveling along a classical trajectory defined by the motion of the lattice.  The $\varepsilon_n(0,\tilde{\Omega})$ and $p_n(\alpha,\tilde{\Omega})$ terms represent the band structure of the lattice, with the $p_n(\alpha,\tilde{\Omega})$ term accounting for the bands deviating from being flat.  For interferometer geometries in which lattices act as waveguides for the atoms, the net contributions to the phase shift from the $\varepsilon_0(0,\tilde{\Omega})$ and $p_0(\alpha,\tilde{\Omega})$ terms in the ground state energy and from corrections to the adiabatic approximation are often negligible, as explained in the following sections.  In this case, the phase difference between the two arms of the interferometer is given by (assuming that the two arms arrive at the same endpoint):

\begin{eqnarray} \label{eqn:Feynman}
\nonumber \phi_1-\phi_2&=&- \frac{1}{\hbar}\left[ \int_{0}^{T} \varepsilon_0^{\text{arm1}}(t) d t-\int_{0}^{T} \varepsilon_0^{\text{arm2}}(t) d t\right] \\
\nonumber &=&\frac{1}{\hbar}\left[ \int_{0}^{T} \frac{1}{2}mv_{\text{Lattice}}^{\text{arm1}}(t)^2d t-\int_{0}^{T} \frac{1}{2}mv_{\text{Lattice}}^{\text{arm2}}(t)^2 d t \right] \\
\end{eqnarray}

\noindent where $T$ is the time elapsed during the interferometer sequence.

Observe that this expression for the phase difference can be obtained by assuming that the lattice potential acts as a constraint that forces the atoms in each arm to traverse the classical path traveled by the lattice guiding that arm.  In this case the phase shift is just the difference of the respective action integrals over the two classical paths, as we would expect from the Feynman path integral formulation of quantum mechanics \cite{ref:feynmanhibbs}.  Since the lattice is the only potential in the freely falling and dressed state frames, the action integrals yield Eq. (\ref{eqn:Feynman}) (for an insightful treatment of the applications of path integrals in atom interferometry, we refer the reader to \cite{ref:storey}).  The terms that we have neglected in Eq. (\ref{eqn:Feynman}) embody corrections to the simple picture of the lattice as a force of constraint arising from the quantum nature of the motion (e.g., a small portion of the population leaving the ground state of the lattice), which can sometimes be important.  However, our simple picture provides physical intuition into the lattice phase shift and is often sufficient to derive quantitative results.

To summarize, the eigenvalues of the lattice consist of a kinetic energy term, a term that depends only on the lattice depth, and a term that is periodic in $\alpha$, and the eigenvectors transform under a simple shift operation when $\alpha$ is changed by $2 m$ for integer $m$.  These properties are a direct result of Bloch's theorem.  The symmetries that we have discussed allow us to conclude that if, for a given $\tilde{\Omega}$, we know the eigenvalues and eigenvectors of the Hamiltonian for all $\alpha$ within any range $[\alpha_0, \alpha_0 + 2]$, we can subsequently determine the eigenvalues and eigenvectors for arbitrary $\alpha$.  This result will prove to be useful from a computational standpoint, since the dynamics of the system are completely described by the solution within a finite range of $\alpha$.

\section{Calculating Corrections to the Adiabatic Approximation Using the Method of Perturbative Adiabatic Expansion}

\noindent

We now present the method of perturbative adiabatic expansion \cite{ref:braggtheory} to determine corrections of arbitrary order to the adiabatic approximation.  We note that the particular adiabatic approximation that we correct here refers to the adiabatic evolution of the ground state of the dressed state Hamiltonian, rather than the adiabatic elimination of the excited state during the Raman process, which is treated in \cite{ref:braggtheory}.  The corrections we consider will always be present to some extent, since lattice depth and velocity ramps occurring over a finite time can never be perfectly adiabatic.  In addition, non-adiabatic corrections can be caused by perturbations arising from laser frequency noise and amplitude noise.  Although our analytical and numerical computations indicate that the contribution of non-adiabatic corrections to the overall phase shift will be highly suppressed for many interferometer geometries, it is important to have a generalized framework with which to treat these corrections in order to determine when they are important and to precisely calculate them when necessary.

During the course of this derivation, it is convenient to parameterize the Hamiltonian, eigenvalues, and eigenvectors using the single variable $t$ rather than the two variables $\alpha(t)$ and $\tilde{\Omega}(t)$.  Note that much of our discussion will follow a similar outline as the proof of the adiabatic theorem in \cite{ref:griffiths}.  A more detailed version of the derivation presented here can be found in \cite{ref:kovachy}.

For all times $t$, we can express any state vector $\vec{\Psi}(t)$ in Hilbert space as a linear combination of the instantaneous eigenvectors $\vec{\Psi}_n(t)$ of the dressed state Hamiltonian, where in general the coefficients of each eigenvector can vary in time.  The instantaneous eigenvectors satisfy the relation $H(t) \vec{\Psi}_n(t) = \varepsilon_n(t) \vec{\Psi}_n(t)$.  Choosing coefficients with a phase $\varphi_n(t) \equiv -\frac{1}{\hbar} \int_{t_0}^{t} \varepsilon_n(t^{\prime}) dt^{\prime}$ factored out, we can write:

\begin{equation} \label{eqn:stateexpansion}
\vec{\Psi}(t) = \sum_{n=0}^{\infty} b_n(t) e^{i \varphi_n(t)} \vec{\Psi}_n(t)
\end{equation}

\noindent  To simplify matters further, we choose the phase of $\vec{\Psi}_n(t)$ so that each element of $\vec{\Psi}_n(t)$ is real for all $t$ and varies continuously with $t$, which we can do because the particular Hamiltonian matrix $H(t)$ we consider is a real-valued, Hermitian matrix.  

When applied to Eq. (\ref{eqn:stateexpansion}), the Schrodinger equation gives us the relation:

\begin{equation} \label{eqn:coefderiv}
\dot{b}_j(t) =  -\sum_{n=0}^{\infty}  b_n(t) \vec{\Psi}^{\dag}_j(t) \frac{\partial \vec{\Psi}_n(t)}{\partial t} e^{i \left[\varphi_n(t) - \varphi_j(t) \right ]}
\end{equation}

\noindent  We note that the inner products $\vec{\Psi}^{\dag}_j(t) \frac{\partial \vec{\Psi}_j(t)}{\partial t}$ are zero for the case of a linear external potential, as verified numerically, and we will thus drop them from the sum.  These inner products are closely related to the Berry's phase, because integrating $\vec{\Psi}^{\dag}_j(t) \frac{\partial \vec{\Psi}_j(t)}{\partial t}$ with respect to time gives the Berry's phase for the $j$th eigenvector.  Eq. (\ref{eqn:coefderiv}) provides us with a relation which we could directly use to perform adiabatic expansion.  But to see more clearly how the adiabatic expansion series relates to the rate at which the Hamiltonian changes in time, we will express $\dot{b}_j(t)$ in a more transparent form.  As long as $\varepsilon_n(t)-\varepsilon_j(t) \neq 0$, we can express $\vec{\Psi}^{\dag}_j(t) \frac{\partial \vec{\Psi}_n(t)}{\partial t}$ in terms of a matrix element of $\dot{H}(t)$ and an energy difference \cite{ref:kovachy}:

\begin{equation} \label{eqn:matrixelmnt}
\vec{\Psi}^{\dag}_j(t) \frac{\partial \vec{\Psi}_n(t)}{\partial t} = \frac{\vec{\Psi}^{\dag}_j(t) \dot{H}(t) \vec{\Psi}_n(t)}{\varepsilon_n(t)-\varepsilon_j(t)}
\end{equation}

\noindent  We can therefore write Eq. (\ref{eqn:coefderiv}) as:

\begin{equation} \label{eqn:non-adiabaticexpansion}
\dot{b}_j(t) =  -\left(\sum_{n \in S_{D}(t)} b_n(t) \vec{\Psi}^{\dag}_j(t) \frac{\partial \vec{\Psi}_n(t)}{\partial t} e^{i \left[\varphi_n(t) - \varphi_j(t) \right ]} \right)  - \left(\sum_{n \in S_{ND}(t)} b_n(t) \frac{\vec{\Psi}^{\dag}_j(t) \dot{H}(t) \vec{\Psi}_n(t)}{\varepsilon_n(t)-\varepsilon_j(t)} e^{i \left[\varphi_n(t) - \varphi_j(t) \right ]}\right)
\end{equation}

\noindent where $S_{D}(t)$ is the set of all $n$ such that $n \neq j$ and  $\varepsilon_n(t) = \varepsilon_j(t)$ and $S_{ND}(t)$ is the set of all $n$ such that $\varepsilon_n(t) \neq \varepsilon_j(t)$.

Eq. (\ref{eqn:non-adiabaticexpansion}) illuminates the rationale behind the adiabatic approximation.  Under the adiabatic approximation, we assume that $H(t)$ and hence also its eigenvectors vary slowly enough in time so that the conditions $\left |\vec{\Psi}^{\dag}_j(t) \dot{H}(t) \vec{\Psi}_n(t) \right | \ll \frac{(\varepsilon_n(t)-\varepsilon_j(t))^2}{\hbar}$ (for $\varepsilon_n(t) \neq \varepsilon_j(t)$) and $\left | \vec{\Psi}^{\dag}_j(t) \frac{\partial \vec{\Psi}_n(t)}{\partial t} \Delta t\right | \ll 1$ (for $\varepsilon_n(t) = \varepsilon_j(t)$ and where $\Delta t$ is the time scale of the approximation) hold.  The righthand side of Eq. (\ref{eqn:non-adiabaticexpansion}) can therefore be approximated as zero.  Then, all the coefficients $b_j(t)$ are constant in time.

To compute higher order corrections, we employ the method of adiabatic expansion,  which mathematically follows in the spirit of the Born approximation.  To zeroth order, we take the coefficients $b_j(t)$ to be constant as dictated by the adiabatic approximation.  To obtain the first order corrections to these coefficients, we substitute the constant zeroth order coefficients $b^{(0)}_j$ into the righthand side of Eq. (\ref{eqn:coefderiv}) with $n \neq j$ and integrate to find the first order coefficients $b^{(1)}_j(t)$.  We can repeat this process recursively, substituting the coefficients $b^{(1)}_j(t)$ into Eq. (\ref{eqn:coefderiv}) to calculate the coefficients $b^{(2)}_j(t)$ and continuing until we know the coefficients to the necessary precision.  This method provides a way to construct a series expansion for each $b_j(t)$.  We adopt a matrix notation for the terms in this series to facilitate the discussion.  The first order correction to $b_j(t)$ consists of contributions from each nonzero $b^{(0)}_n$ where $n \neq j$, and we depict the contribution from  $b^{(0)}_n$ as $C_{n \rightarrow j}$.  So to first order, we can write:

\begin{equation}
b^{(1)}_j(t)= b^{(0)}_j + \sum_{n \neq j} C_{n \rightarrow j}
\end{equation}

\noindent where:

\begin{equation} \label{eqn:correction1}
C_{n \rightarrow j} \equiv -\int_{t_0}^{t} b^{(0)}_n \vec{\Psi}^{\dag}_j(t^{\prime}) \frac{\partial \vec{\Psi}_n(t^{\prime})}{\partial t^{\prime}} e^{i \left[\varphi_n(t^{\prime}) - \varphi_j(t^{\prime}) \right ]} dt^{\prime}
\end{equation}

\noindent  The second order solution for $b_j(t)$ will then be:

\begin{equation} \label{eqn:secondorder}
b^{(2)}_j(t) = b^{(0)}_j - \sum_{n \neq j}\int_{t_0}^{t} b^{(1)}_n(t^{\prime}) \vec{\Psi}^{\dag}_j(t^{\prime}) \frac{\partial \vec{\Psi}_n(t^{\prime})}{\partial t^{\prime}} e^{i \left[\varphi_n(t^{\prime}) - \varphi_j(t^{\prime}) \right ]} dt^{\prime} = b^{(0)}_j+ \sum_{n \neq j} C_{n \rightarrow j} +  \sum_{n \neq j} \sum_{m \neq n} C_{m \rightarrow n \rightarrow j}
\end{equation}

\noindent where:

\begin{equation}  \label{eqn:factor2}
C_{m \rightarrow n \rightarrow j} \equiv -\int_{t_0}^{t} C_{m \rightarrow n} \vec{\Psi}^{\dag}_j(t^{\prime}) \frac{\partial \vec{\Psi}_n(t^{\prime})}{\partial t^{\prime}} e^{i \left[\varphi_n(t^{\prime}) - \varphi_j(t^{\prime}) \right ]} dt^{\prime}
\end{equation}

\noindent under the implicit assumption that the time variable associated with $C_{m \rightarrow n}$ is appropriate for the context in which it appears.  The calculation of corrections of higher order is discussed in Appendix D.

To find the eigenvectors and eigenvalues that we need to calculate the terms that make a non-negligible contribution to the expansion, we must approximate the infinite dimensional Hamiltonian matrix as a finite dimensional truncated matrix.  At the end of Sec. III., we concluded that the problem of determining the eigenvalues and eigenvectors for all $\alpha$ reduces to finding the eigenvalues and eigenvectors for a range $\alpha \in [\alpha_0, \alpha_0+2]$ for arbitrary $\alpha_0$.  In addition, it suffices to calculate the inner products $\vec{\Psi}^{\dag}_j(t) \frac{\partial \vec{\Psi}_n(t)}{\partial t}$ just in this range of $\alpha$, which follows from the symmetry $\vec{\Psi}^{\dag}_j(\alpha + 2m,\tilde{\Omega}) \frac{\partial \vec{\Psi}_n(\alpha + 2m,\tilde{\Omega})}{\partial t}  = \vec{\Psi}^{\dag}_j(\alpha,\tilde{\Omega}) \frac{\partial \vec{\Psi}_n(\alpha,\tilde{\Omega})}{\partial t}$ for integer $m$ \cite{ref:kovachy}.  To make the calculation less cumbersome, we can look at the range
$\alpha \in [-1,1]$.  For $\tilde{\Omega}$ not too large and $\alpha$ in this range, the eigenvectors with lower energies are populated almost entirely by momentum eigenstates $\left | 2m \hbar k \right >$ with relatively small $|m|$.  This is the case because for $\alpha$ in the range, the diagonal elements of the Hamiltonian will be smallest for values of $m$ close to zero.  We note that for $\tilde{\Omega} =0$, the diagonal elements are the eigenvalues.  In the limit of $\tilde{\Omega} \rightarrow 0$, each eigenvector will consist of only a single momentum eigenstate, where in general eigenvectors corresponding to momentum eigenstates with $m$ closer to zero will have lower eigenvalues. Increasing $\tilde{\Omega}$ will allow the lower eigenvectors to spread out in momentum space to a certain extent, but this will not change the fact that the lower eigenvectors will be linear combinations of momentum eigenstates corresponding to smaller values of $|m|$.  As discussed previously, the eigenvectors we care about for calculational purposes will be those with eigenvalues closer to the ground state eigenvalue.  We can thus consider a truncated $(2n+1) \times (2n+1)$ Hamiltonian matrix centered around $m=0$, where we choose $n$ to be large enough so that for the particular dynamics being described, a sufficient number of eigenvectors and eigenvalues can be calculated.

\section{An Example of Perturbative Adiabatic Expansion:  Calculating the Non-Adiabatic Correction to the Phase Shift Evolved During a Lattice Beam Splitter}

\noindent

We illustrate the above method by calculating phase corrections to a lattice beam splitter.  In this example, we consider the case where two optical lattices of the same depth but different accelerations are used to separate the two arms of the interferometer (after an initial momentum space splitting is achieved through Bragg diffraction).  We note that the analysis here is equally applicable to the situation where two separate optical lattice waveguides are used to address the two arms of the interferometer, an example of which is illustrated in Fig. \ref{fig:conj}.  To calculate the phase shift for applications in precision measurement, we need to determine the non-adiabatic correction to the phase difference between the two arms that accrues during the beam splitter.  In practice, we do this by first calculating corrections to the ground state coefficient $b_0(t)$ and then evaluating how these corrections affect the phase difference between the arms.  We note that the dominant contribution to the phase difference will come from the zeroth order term as given in Eq. (\ref{eqn:Feynman}).

For this example, we consider the situation shown in Fig. \ref{fig:latticeaccel} ,where the interaction of the atoms with the lattice is divided into three distinct parts.  From $t = 0$ to $t = T_{\text{ramp}}$ we ramp up the lattice, from $t = T_{\text{ramp}}$ to $t = T_{\text{ramp}} + T_{\text{accel}}$ we accelerate the lattice, and from $t = T_{\text{ramp}} + T_{\text{accel}}$ to $t = 2 T_{\text{ramp}} + T_{\text{accel}} \equiv T_{\text{final}}$ we ramp down the lattice.  For the sake of simplicity, the ramps are chosen to be symmetric so that the lattice depth decrease ramp is the time reversed lattice depth increase ramp.

We assume that initially all of the population is in the ground state, so that $b^{(0)}_j = \delta_{0j}$, and we make the lattice depth and velocity ramps adiabatic enough so that almost all of the population remains in the ground state.  In order to find the non-adiabatic correction to the phase shift, we determine the non-adiabatic correction to the phase of the ground state for each arm.

Since to lowest order only the ground state is populated, the leading corrections to $b_0(t)$ will come at second order.  Using the formalism in Eq. (\ref{eqn:secondorder}), the second order correction will be:

\begin{equation}
b^{(2)}_0(t) - b^{(0)}_0= \sum_{n \neq 0} C_{0 \rightarrow n \rightarrow 0}(t)
\end{equation}

\noindent The largest contribution comes from $C_{0 \rightarrow 1 \rightarrow 0}(t)$, and it is this term on which we focus.  Because the ground state is non-degenerate, Eqs. (\ref{eqn:non-adiabaticexpansion}) and (\ref{eqn:factor2}) give us:

\begin{eqnarray} \label{eqn:correctionintegrals}
\nonumber C_{0 \rightarrow 1 \rightarrow 0 }(t)&=& -\int_{0}^{t} dt_1C_{0 \rightarrow 1}(t_1) \frac{\vec{\Psi}^{\dag}_0(t_1) \dot{H}(t_1) \vec{\Psi}_1(t_1)}{\varepsilon_1(t_1)-\varepsilon_0(t_1)} e^{i \left[\varphi_1(t_1) - \varphi_0(t_1) \right ]}   \\
\nonumber &=& -\int_{0}^{t} dt_1 \left(-\int_{0}^{t_1} dt_2 \frac{M_{10}(t_2)}{\Delta \varepsilon_{10}(t_2)} e^{-\frac{i}{\hbar} \int_{0}^{t_2} \Delta \varepsilon_{10}(t_3) dt_3} \right) \frac{M_{10}(t_1)}{- \Delta \varepsilon_{10}(t_1)} e^{\frac{i}{\hbar} \int_{0}^{t_1} \Delta \varepsilon_{10}(t_3) dt_3} \\
\end{eqnarray}

\noindent where we define $M_{10}(t) \equiv \vec{\Psi}^{\dag}_1(t) \dot{H}(t) \vec{\Psi}_0(t) = \vec{\Psi}^{\dag}_0(t) \dot{H}(t) \vec{\Psi}_1(t)$ (note that the two matrix elements are equal because we choose the eigenvectors to be real) and $\Delta \varepsilon_{10}(t) \equiv \varepsilon_0(t)-\varepsilon_1(t)$.

We examine the ultimate contribution of $C_{0 \rightarrow 1 \rightarrow 0}(T_{\text{final}})$ to $b^{(2)}_0(T_{\text{final}})$.  Since during the ramp up and ramp down stages $M_{10}(t)$ depends only on $\dot{\tilde{\Omega}}$ but not on $\dot{\alpha}$, some portions of $C_{0 \rightarrow 1 \rightarrow 0}(T_{\text{final}})$ will be common to both arms of the interferometer, because we assume that the lattice interaction processes for the two arms differ only in the magnitude of the lattice acceleration.  We denote these common terms as $g_{\text{ramp}}$.  The remaining terms depend on the lattice acceleration and will thus differ between the arms.  There will be a term $g_{\text{mixed}}$ that depends both on $\dot{\tilde{\Omega}}$ and $\dot{\alpha}$.  However, it can be shown that under the assumption that the lattice depth decrease ramp is the time reversed lattice depth increase ramp, this term is zero \cite{ref:kovachy}.  Finally, there will be a term $g_{\text{accel}}$ that depends quadratically on $\dot{\alpha}$ and is not explicitly dependent on $\dot{\tilde{\Omega}}$.  We note that $g_{\text{accel}}$ implicitly depends on the maximum lattice depth $\tilde{\Omega}_{\text{max}}$, which can be seen by the fact that $\tilde{\Omega}_{\text{max}}$ affects what value the energy gap $\Delta \varepsilon_{10}$ takes on during the acceleration stage (a deeper lattice leads to a larger energy gap).  We can thus write:

\begin{equation}
b^{(2)}_0(T_{\text{final}}) \approx 1+ g_{\text{ramp}} +g_{\text{accel}}
\end{equation}

In calculating $g_{\text{accel}}$, it is useful to note that $M_{10}(t)$ takes on a convenient form during the acceleration stage.  Recalling the form of the Hamiltonian matrix from Eq. (\ref{eqn:hamiltonianelements}), we observe that $\dot{H}(t)$ will be a diagonal matrix with matrix elements $\dot{H}_{mn}(t) = -4E_r n \dot{\alpha}(t) \delta_{mn}$.  We can thus write $M_{10}(t) = \dot{\alpha}(t) A_{10}(t)$, where $A_{10}(t) \equiv -4 E_r \sum_{n} n \left(\vec{\Psi}_1(t)\right)_n \left( \vec{\Psi}_0(t)\right)_n$ is a weighted dot product.

In order to more clearly illuminate the general points we are illustrating with this example, we make the simplifying assumption that $\dot{\alpha}(t)$ is constant throughout the acceleration stage.  Moreover, we assume that the lattice is deep enough so that $A_{10}(t)$ and $\Delta \varepsilon_{10}(t)$ are also constant during the acceleration stage, which is an accurate approximation for typical experimental situations.  During the acceleration stage, we respectively denote these constant quantities as $\dot{\alpha}$, $A_{10}$, and $\Delta \varepsilon_{10_{\text{accel}}}$.  Note that these assumptions, along with the assumption of mirror symmetry between the ramp up and ramp down stages, are certainly not necessary to carry out the calculation.  They only serve to make the final result take a particularly simple form that provides physical insight into the process.  In the absence of these assumptions, the calculation will be only slightly more complicated and can easily be performed.  We note in particular that the mirror symmetry assumption is not stringent, for even when this symmetry is largely violated, the $g_{\text{mixed}}$ term is typically an order of magnitude or more smaller than the $g_{\text{accel}}$ term, as we verify by estimating the relevant integrals in Eq. (\ref{eqn:correctionintegrals}).  If needed, $g_{\text{mixed}}$ can be calculated by evaluating these integrals.  In addition, we note that the treatment given in this example can readily be generalized to the case where the lattice depth and velocity are changed simultaneously.  Performing the necessary integrals, we find that:

\begin{equation}
g_{\text{accel}} = -i \hbar \frac{\dot{\alpha}^2 A_{10}^2}{(\Delta \varepsilon_{10_{\text{accel}}})^3} T_{\text{accel}} + \hbar^2 \frac{\dot{\alpha}^2 A_{10}^2}{(\Delta \varepsilon_{10_{\text{accel}}})^4} \left[e^{\frac{i}{\hbar}\Delta \varepsilon_{10_{\text{accel}}} T_{\text{accel}}} - 1 \right]
\end{equation}

\noindent  We note that for $\left | g_{\text{ramp}} +g_{\text{accel}} \right | \ll 1$, we can solve the problem by employing adiabatic expansion over a single time interval.  However, there may be times when we must divide the problem into multiple parts, as discussed in Appendix D.  For typical experimental parameters, the term proportional to $T_{\text{accel}}$ in $g_{\text{accel}}$ will dominate both the second term in $g_{\text{accel}}$ and the $g_{\text{ramp}}$ term.  We have verified that the $g_{\text{ramp}}$ term (which embodies the non-adiabatic loading of the lattice) is typically much smaller than the second term in $g_{\text{accel}}$ by estimating the integrals in Eq. (\ref{eqn:correctionintegrals}) that correspond to $g_{\text{ramp}}$ and by checking these estimates numerically.  Thus, we can express the condition $\left | g_{\text{ramp}} +g_{\text{accel}} \right | \ll 1$ as:

\begin{equation}
\left | \hbar \frac{\dot{\alpha}^2 A_{10}^2}{(\Delta \varepsilon_{10_{\text{accel}}})^3} T_{\text{accel}} \right |\ll1
\end{equation}

\begin{figure}
\includegraphics[width=7.00in]{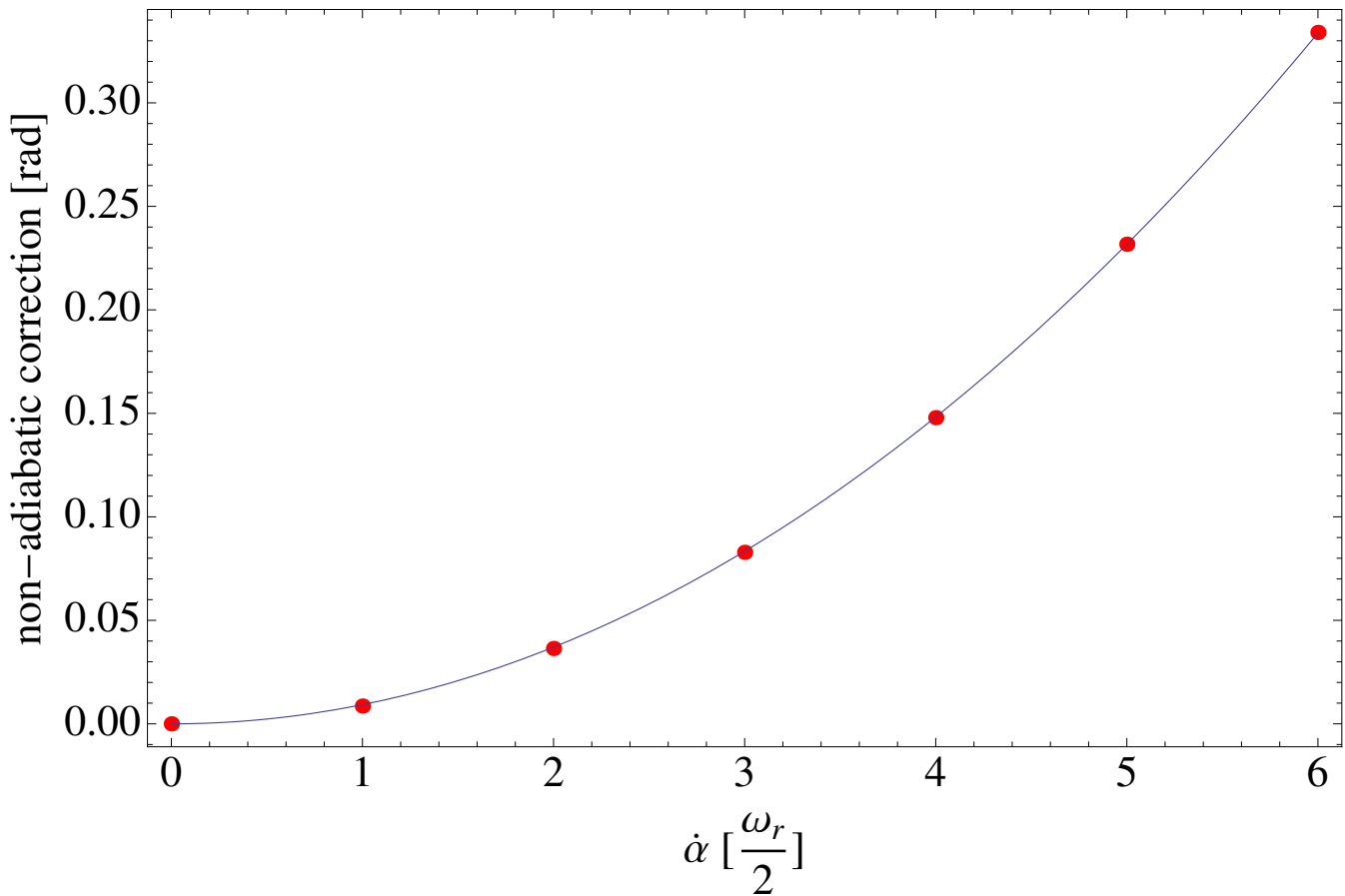}
\caption{(color online) Non-adiabatic correction to the phase difference between an arm that is accelerated by an experimentally plausible lattice beam splitter and an arm that is not accelerated (which is in our notation $\delta \phi (\dot{\alpha})$), as calculated using a simplified adiabatic expansion method in which we keep only leading terms versus numerical simulations of the Schrodinger equation.  The curve represents the prediction made by the adiabatic expansion method, while the red dots represent the numerical results.  The lattice depth ramps and acceleration time are identical to those of the acceleration sequence shown in Fig. \ref{fig:latticeaccel}, with an acceleration time of $16(4 \omega_r)^{-1}$ and with $\tilde{\Omega}_{\text{max}} = \frac{19.5}{8}$.  In addition to the leading second order term, we keep the leading fourth order correction term in $g_{accel}$, $-\hbar^2 \frac{\dot{\alpha}^4 A_{10}^4 T_{accel}^2}{2\left(\Delta \varepsilon_{10_{accel}}\right)^6}$, calculated in \cite{ref:kovachy}, which becomes significant for larger accelerations (e.g., for an acceleration of $6 \left(\frac{\omega_r}{2}\right) v_r$, this term is smaller than the leading second order correction by a factor of $\sim 10$).  We neglect corrections arising from the term $g_{\text{ramp}}$, for these corrections are common to both arms of the interferometer to lowest order.  Even the simple approximation used to obtain the curve agrees remarkably well with the simulations (with an rms deviation of $4 \times 10^{-4}$ radians), and we note that we could easily improve this approximation by including more terms in the adiabatic expansion series.  As expected, we observe that the correction scales quadratically with acceleration.}
\label{fig:correction}
\end{figure}

\noindent  This condition will often hold, since in many experimentally relevant cases the acceleration time or acceleration will be sufficiently small.

We now show how to determine the correction to the phase shift between the two arms arising from the non-adiabatic correction $g_{\text{ramp}} +g_{\text{accel}}$ in the case where this correction is small.  As we recall from Eq. (\ref{eqn:stateexpansion}), the coefficient of the ground state eigenvector at time $T_{\text{final}}$ is $b_0(T_{\text{final}}) e^{i \varphi_0(T_{\text{final}})} \approx \left(1 + g_{\text{ramp}} +g_{\text{accel}} \right) e^{i \varphi_0(T_{\text{final}})}$.  Now, note that the argument of any complex number $z = |z| e^{i \arg(z)}$ can be written as $\arg(z) = \frac{1}{2 i}\ln\left[\frac{z}{z^*}\right]$.  The non-adiabatic correction to the phase of the coefficient of the ground state eigenvector for an arm with acceleration $\dot{\alpha}$ is thus:

\begin{eqnarray} \label{eqn:phasecorrection}
\nonumber \phi_{\text{correction}}(\dot{\alpha}) &=& \frac{1}{2 i} \ln\left[\frac{\left(1 +\left|g_{\text{ramp}}\right| e^{i\arg\left(g_{\text{ramp}}\right)} + \left|g_{\text{accel}}\right| e^{i\arg\left(g_{\text{accel}}\right)} \right)}{\left(1 +\left|g_{\text{ramp}}\right| e^{-i\arg\left(g_{\text{ramp}}\right)} + \left|g_{\text{accel}}\right| e^{-i\arg\left(g_{\text{accel}}\right)} \right)}\right] \\
\nonumber \\
\nonumber  &\approx& \left|g_{\text{ramp}}\right| \sin \left(\arg\left[g_{\text{ramp}}\right]\right) +\left|g_{\text{accel}}\right| \sin \left(\arg\left[g_{\text{accel}}\right]\right) \\
\end{eqnarray}

\noindent where we have Taylor expanded to first order in small quantities.  To calculate the correction to the phase shift between two arms, we take the difference $\phi_{\text{correction}}(\dot{\alpha}_{\text{arm1}})-\phi_{\text{correction}}(\dot{\alpha}_{\text{arm2}})$.  The $ \left|g_{\text{ramp}}\right| \sin \left(\arg\left[g_{\text{ramp}}\right]\right) $ term is common to both arms.  The non-adiabatic correction to the phase difference between an arm with acceleration $\dot{\alpha}$ and an unaccelerated arm is:

\begin{equation}  \label{eqn:phasecorrectionref}
\delta \phi (\dot{\alpha}) \equiv \phi_{\text{correction}}(\dot{\alpha}) - \phi_{\text{correction}}(0) \approx \left|g_{\text{accel}}\right| \sin \left(\arg\left[g_{\text{accel}}\right]\right)
\end{equation}

Eqs. (\ref{eqn:phasecorrection}) and (\ref{eqn:phasecorrectionref}) provide us with a means to calculate the leading correction to the phase shift, and comparison with numerical results for a variety of experimentally conceivable lattice depth ramps shows excellent agreement.  A comparison of the adiabatic expansion method with numerical calculations is illustrated in Fig. \ref{fig:correction}.  In particular, Fig. \ref{fig:correction} shows $\delta \phi (\dot{\alpha})$ as a function of $\dot{\alpha}$ for an acceleration time of $16(4 \omega_r)^{-1}$ with $\tilde{\Omega}_{\text{max}} = \frac{19.5}{8}$.  The numerical values shown in the figure come from a numerical simulation of the Schrodinger equation.  The non-adiabatic correction to the phase shift between the two arms of the interferometer for arbitrary arm acceleration differences is $\phi_{\text{correction}}(\dot{\alpha}_{\text{arm1}})-\phi_{\text{correction}}(\dot{\alpha}_{\text{arm2}})=\delta \phi(\dot{\alpha}_{\text{arm1}}) - \delta \phi(\dot{\alpha}_{\text{arm2}})$.  Note that the leading order results depicted in Fig. \ref{fig:correction} will often be sufficient, but it may sometimes be necessary to calculate higher order corrections, examples of which can be found in \cite{ref:kovachy}.

In an experimental implementation, the two arms of the beam splitter are addressed by two different lattices.  Therefore, any imbalance in the depths of the two lattices will lead to a phase error between the arms.  Where the intensities of the two beams forming a given lattice are $I_a$ and $I_b$, we note that the lattice depth is proportional to the product $\sqrt{I_a} \times \sqrt{I_b}$, meaning that intensity imbalances lead to lattice depth imbalances.  We calculate that once a lattice is ramped up, the $\varepsilon_0(0,\tilde{\Omega})$ term in the expression for the lowest eigenvalue of the dressed state Hamiltonian from Eq. (\ref{eqn:evalue}) is much larger than the $p_0(\alpha,\tilde{\Omega})$ term.  Thus, for an interaction lasting from time $t_1$ to time $t_2$, the dominant contribution to the phase error, which we denote as $\phi_{\text{balance}}$, arising from the imbalance in the lattice depths is:

\begin{equation} \label{eqn:balance}
\ \phi_{\text{balance}} =  - \frac{1}{\hbar}\int_{t_1}^{t_2} \left[ \varepsilon_0(0,\tilde{\Omega}_{\text{arm1}}) -\varepsilon_0(0,\tilde{\Omega}_{\text{arm2}}) \right] d t
\end{equation}

\noindent where we note that $\varepsilon_0(0,\tilde{\Omega})$ can be calculated using the truncated matrix approximation discussed in Sec. IV.  In order to put Eq. (\ref{eqn:balance}) in a more convenient form for making order of magnitude estimates, we use the fact that for $|\tilde{\Omega}| > 0.25$, $\varepsilon_0(0,\tilde{\Omega}) \sim - 8 E_r | \tilde{\Omega} |$.  This result is verified by direct comparison with the values of $\varepsilon_0(0,\tilde{\Omega})$ obtained with the truncated matrix approximation.  Recalling that $\tilde{\Omega} \equiv \frac{\Omega}{8 \omega_r}$, it is convenient to rephrase this statement in terms of $\Omega$ as follows:  for $|\Omega|>2 \omega_r$, $\varepsilon_0(0,\Omega) \sim - \hbar | \Omega |$.  This range of $|\Omega|$ is of  considerable experimental interest and is amenable to simple approximation.  However, if needed, smaller lattice depths can be treated using the general relation given in Eq. (\ref{eqn:balance}) and the truncated matrix approximation.  Substituting the above result into Eq. (\ref{eqn:balance}), we obtain:

\begin{equation} \label{eqn:balanceapprox}
\ \phi_{\text{balance}} \sim \int_{t_1}^{t_2} \left( | \Omega_{\text{arm1}} | -|\Omega_{\text{arm2}}| \right) d t
=  \left (\left< | \Omega_{\text{arm1}} | \right> - \left< | \Omega_{\text{arm2}} | \right>\right) (t_2 - t_1)
\end{equation}

\noindent where $\left< | \Omega_{\text{arm1}} | \right>$ and $\left< | \Omega_{\text{arm2}} | \right>$ respectively denote the average values of $ |\Omega_{\text{arm1}}|$ and $ |\Omega_{\text{arm2}}|$ between time $t_1$ and time $t_2$.  Note that fluctuations in the difference between $ |\Omega_{\text{arm1}}|$ and $|\Omega_{\text{arm2}}|$ that occur at frequencies that are large with respect to the beam splitter time $t_2 - t_1$ will largely average out, suppressing their net effect on the phase shift.  Furthermore, in many geometries, arm 1 will be addressed by one pair of laser beams, that we call lattice A, during the splitting of the arms and then will be addressed by a second pair of laser beams, that we call lattice B, during the recombination of the arms.  Conversely, arm 2 will be addressed by lattice B during the splitting stage and by lattice A during the recombination stage.  In such a geometry, the effect on the phase shift of a constant offset in depth between lattice A and lattice B will cancel, and the effect on the phase shift of fluctuations in the depth difference between lattice A and lattice B that occur at low frequencies with respect to the time scale of the interferometer sequence will be highly suppressed.  Also, as we discuss in the following section, we will often be interested in the difference in the phase shifts of two interferometers in a differential configuration.  If both of the interferometers remain well within the Rayleigh ranges of the laser beams so that beam divergence is a small effect, any lattice depth imbalance will be largely common to the two interferometers, suppressing its effect on the phase shift difference by orders of magnitude.

\section{Applications:  Atom Interferometers Using Optical Lattices as Waveguides}

Light-pulse atom interferometer geometries have had tremendous success in performing many types of high-precision measurements.  However, in many cases, we would like to be able to push the capabilities of atom interferometry by making more precise measurements using spatially compact interferometers.  Atom interferometers that use optical lattices as waveguides for the atoms offer the potential to make such measurements attainable.  In such a scheme, we can use an initial beam splitter composed of multiple Bragg pulses, a multi-photon Bragg pulse,  or a hybrid Bragg pulse/lattice acceleration scheme as described in \cite{ref:denschlag,ref:clade,ref:muellerlattice} to split the arms of the interferometer in momentum space.  We can then control each arm independently with an optical lattice.  We will once again use Bragg pulses during the $\pi$-pulse and final $\frac{\pi}{2}$-pulse stages of the interferometer sequence, with lattices acting as waveguides between these stages.  The preceding analysis has developed the theoretical machinery for calculating phase shifts for these lattice interferometers.  We now examine several of the most promising applications of lattice interferometers.

\subsection{Gravimetry and gravity gradiometry}

Lattice interferometers can be used to make extremely precise measurements of the local gravitational acceleration $g$.  We proceed to calculate the phase shift for a lattice gravimeter.  It is essential to note that whenever the two arms are addressed by different lattices they will be in different dressed state frames (where we recall that a dressed state frame is defined by the velocity of the corresponding lattice and by the distance that the lattice has traveled since the beginning of the interferometer sequence).  Let the velocities in the lab frame of the two lattices be denoted as $v_{\text{Lab}}^{\text{arm1}}(t)$ and $v_{\text{Lab}}^{\text{arm2}}(t)$, respectively.  The lattice velocities for the two arms in their respective dressed state frames will thus be $v_{\text{Lattice}}^{\text{arm1}}(t)=v_{\text{Lab}}^{\text{arm1}}(t)+gt - v_0$ and $v_{\text{Lattice}}^{\text{arm2}}(t)=v_{\text{Lab}}^{\text{arm2}}(t)+gt - v_0$.  We note that $v_0$ is the velocity of the atom before the initial Bragg diffraction that splits the arms in momentum space, so that the two arms have different momenta after the Bragg diffraction.  During the lattice loading period, the two lattices must be resonant (as described in Sec. III and in Appendix C) with the respective portions of the atomic wavefunction that they are addressing.  Also, let $\Delta v(t) \equiv v_{\text{Lab}}^{\text{arm1}}(t) - v_{\text{Lab}}^{\text{arm2}}(t)$ be the velocity difference between the two arms and $\Delta d(t) \equiv \int_{0}^{t} \Delta v(t^{\prime}) dt^{\prime}$ be the distance between the two arms.  Where the interferometer sequence lasts for a time $T$, we can derive the phase shift for a lattice gravimeter using Eq. (\ref{eqn:Feynman}).  We note that an additional contribution to the phase shift will arise if the two arms of the interferometer end up in slightly different dressed state frames.  The arms begin in the same dressed state frame, and if $\Delta d(T) = 0$ they will end up in the same dressed state frame, in which case Eq. (\ref{eqn:Feynman}) provides the final say on the phase shift.  If $\Delta d(T)$ differs slightly from zero, the two arms will end up in slightly different dressed state frames.  Since we must compare phases in a common frame, it is convenient to boost both arms into the freely falling frame using the transformation given in Eq. (\ref{eqn:dstransform}) and the mathematical framework discussed in Appendix B.  In addition to the phase difference from Eq. (\ref{eqn:Feynman}), we will then have an additional term in the phase shift equal to $-\frac{1}{\hbar} m(v_f - v_0 + gT) \Delta d(T)$, where $m v_f$ is the lab frame momentum at time $T$ of a particular momentum eigenstate in the atomic wavepacket.  After averaging, it follows from the discussion in \cite{ref:varenna} that $m v_f$ will take on the value of the center of the momentum space wavepacket.  We can combine the contribution from Eq. (\ref{eqn:Feynman}) and the contribution from boosting the two arms to a common frame to calculate the phase shift, where we also include a term $\phi_{\text{Bragg}}$ to embody the net contribution to the phase shift arising from the Bragg pulses:

\begin{eqnarray} \label{eqn:phasederive}
\nonumber \Delta \phi&=&\frac{1}{\hbar}\left[ \int_{0}^{T} \frac{1}{2}m(v_{\text{Lab}}^{\text{arm1}}(t)+gt - v_0)^2d t-\int_{0}^{T} \frac{1}{2}m(v_{\text{Lab}}^{\text{arm2}}(t)+gt - v_0)^2 d t \right] -\frac{1}{\hbar} m(v_f - v_0 + gT) \Delta d(T) + \phi_{\text{Bragg}}\\
\nonumber \\
\nonumber &=&\frac{1}{\hbar}\left[ \int_{0}^{T} \frac{1}{2}m(v_{\text{Lab}}^{\text{arm1}}(t)^2-v_{\text{Lab}}^{\text{arm2}}(t)^2)d t+\int_{0}^{T}mg \Delta v(t) t dt - \int_{0}^{T}m v_0 \Delta v(t) dt \right] -\frac{1}{\hbar} m(v_f - v_0 + gT) \Delta d(T) + \phi_{\text{Bragg}}\\
\end{eqnarray}

\noindent Integrating the second term by parts and simplifying yields:

\begin{equation} \label{eqn:phaseshift}
\Delta \phi=- \frac{1}{\hbar}\int_{0}^{T}mg \Delta d(t) dt + \frac{1}{\hbar}\int_{0}^{T} \frac{1}{2}m(v_{\text{Lab}}^{\text{arm1}}(t)^2-v_{\text{Lab}}^{\text{arm2}}(t)^2)d t - \frac{m}{\hbar}v_f \Delta d(T) + \phi_{\text{Bragg}}
\end{equation}

\noindent  When expressed in terms of lab frame quantities, it is apparent that $\Delta \phi$ contains terms corresponding to the propagation phase and the separation phase that typically appear in standard atom interferometer phase shift calculations.  We note that in the dressed state frame, ideal Bragg pulses simply yield contributions in units of $\pm \frac{\pi}{2}$ to the overall phase shift between the arms, and these contributions can easily be made to cancel so that $\phi_{\text{Bragg}} =0$.  However, we note that in some cases, corrections to the simplified picture of an ideal Bragg pulse due to such factors as gravity gradients, finite pulse and detuning effects (which can sometimes lead to a non-negligible propagation phase during the Bragg pulse), phase noise, or population loss may need to be considered.  To avoid unnecessarily complicating our presentation, we will not present these corrections here.  Instead, we emphasize that they are well-understood effects and refer the reader to other sources for further discussion \cite{ref:varenna,ref:braggtheory,ref:antoine}.  Moreover, we note that these effects gain additional suppression for interferometers in a differential configuration so that they will often be below the mrad level \cite{ref:varenna}, as in the case of the gravity gradiometer discussed below.  

In the symmetric case where the velocities of the two arms in the lab frame are either opposite to each other or equal so that $v_{\text{Lab}}^{\text{arm1}}(t)^2-v_{\text{Lab}}^{\text{arm2}}(t)^2 = 0$.  Since we need the two arms of the interferometer to overlap at time $T$, it is convenient for us to choose $\Delta d(T) = 0$.  Thus, the first term in Eq. (\ref{eqn:phaseshift}) will constitute the only contribution to $\Delta \phi$.  However, in an experiment, the parameters in Eq. (\ref{eqn:phaseshift}) will undergo small fluctuations around their desired values from shot to shot, so that the other terms in Eq. (\ref{eqn:phaseshift}) act as a source of noise.  

In order to cancel the effects of this noise, we can adopt a gradiometer setup in which an array of two or more gravimeters interacts with the same lattice beams.  Although fluctuations in $\Delta d(t)$ will still affect phase differences between gravimeters, which take the form $- \frac{1}{\hbar}\int_{0}^{T}m(g_1-g_2) \Delta d(t) dt$, modern phase lock techniques will typically allow us to control the phase differences between the lattice beams well enough so that these effects are smaller than shot noise \cite{ref:muellerlaser}.  When measuring a gravity gradient, the value of $g$ will vary due to the gradient over the range of a single gravimeter.  For linear gradients, we can calculate the phase shift in the presence of a gravity gradient by assuming that the value of $g$ corresponding to the gravimeter is equal to its value at the center of mass position of the atom (see \cite{ref:kovachy} for a rigorous justification of this procedure).  When higher order derivatives of the gravitational field become sizable in comparison to the first derivative, this simple prescription may not suffice, and we can treat the problem perturbatively.  If we want to measure an acceleration as well as a gravitational gradient in a noisy environment in which the fringes of the individual gravimeters are washed out, we can use dissimilar conjugate interferometers whose phase noise is strongly correlated as suggested in \cite{ref:hermann}.  Appropriate statistical methods can then be used to extract the desired signal \cite{ref:stockton}.

The effects on the phase shift of non-adiabatic corrections, lattice depth imbalances, and the finite spread of the atomic wavefunction in momentum space are considered in Sec. IV., Sec. V., and Appendix B.  Based on the analysis in these sections, we conclude that a wide range of experimentally feasible gravimeter geometries exist that contain sufficiently adiabatic lattice depth and velocity ramps and that make use of symmetry in such a way that the net contribution of these corrections to the phase shift will be below the mrad level in a gradiometer configuration.  We note that this paper has developed the mathematical machinery to calculate any such corrections to arbitrary precision if necessary.

When two lattices are used to manipulate the arms of an atom interferometer, one lattice will be on resonance with a given arm, while the other will be highly detuned.  As long as we keep this detuning large enough and/or employ geometries with sufficient symmetry between the arms, the net effect of the off-resonant lattices on the final phase shift can often be made to be smaller than the mrad level in a differential configuration (e.g., a gradiometer).  For arm 1, the detuned lattice will manifest as an additional term in the discrete Hamiltonian, given by:

\begin{equation} \label{eqn:offresmatrix}
\nonumber H_{\text{detuned}} = 4 E_r \left(
\begin{array}{ccccccccccc}
\ddots & \ddots & \ddots & \ddots & \ddots & & &\\
& 0 & -\tilde{\Omega}_{\text{arm2}}e^{-i\beta(t)} & 0 & -\tilde{\Omega}_{\text{arm2}}e^{i\beta(t)}& 0 &  &\\
& & 0 & -\tilde{\Omega}_{\text{arm2}}e^{-i\beta(t)}& 0 & -\tilde{\Omega}_{\text{arm2}}e^{i\beta(t)}& 0 &  & &\\
& & & 0 & -\tilde{\Omega}_{\text{arm2}}e^{-i\beta(t)} & 0 &-\tilde{\Omega}_{\text{arm2}}e^{i\beta(t)}& 0 &  & \\
& & &  & 0 & -\tilde{\Omega}_{\text{arm2}}e^{-i\beta(t)} & 0 &-\tilde{\Omega}_{\text{arm2}}e^{i\beta(t)}& 0 & \\
& & & & &  & \ddots &  \ddots & \ddots & \ddots
\end{array}
\right)
\end{equation}

\noindent where $\beta(t) \equiv - \int_{0}^{t} \left[\alpha^{\text{arm1}}(t^{\prime})-\alpha^{\text{arm2}}(t^{\prime})\right] 4 \omega_rdt^{\prime}=-2 k \Delta d(t)$ and where $\tilde{\Omega}_{\text{arm2}}$ is the lattice depth parameter corresponding to the detuned lattice that addresses arm 2 \cite{ref:malinovsky}.  To obtain the correction to the Hamiltonian for arm 2, which comes from the detuned lattice addressing arm 1, we replace $\beta(t)$ with $-\beta(t)$ and $\tilde{\Omega}_{\text{arm2}}$ with $\tilde{\Omega}_{\text{arm1}}$.  For large detunings, $\beta(t)$ will vary rapidly with time so that the contribution from the detuned Hamiltonian will be small due to the rotating wave approximation.  Corrections arising from the detuned Hamiltonian can be solved for perturbatively using methods such as adiabatic perturbation theory.  But we emphasize again that we can often avoid situations where this will be necessary.  For example, we find from perturbation theory that to avoid population loss due to the off-resonant lattice as described in Sec. III., we should choose the off-resonant lattice to have a velocity that differs from that of the particular arm of the interferometer under consideration by an amount $\Delta v \gg | \tilde{\Omega} | v_r$ (where $\tilde{\Omega}$ is the depth parameter of the off-resonant lattice).  We have verified this result with numerical simulations.  However, in this regime, the off-resonant lattice can still sometimes cause a non-negligible energy shift, which we estimate with perturbation theory.  The energy shift for arm 1 is $\Delta E_{\text{arm1}} \sim \frac{\hbar}{8} \omega_{r}^{-1} (v_r/\Delta v)^{2} (\Omega_{\text{arm2}})^2$.  Analagously, the energy shift for arm 2 is $\Delta E_{\text{arm2}} \sim \frac{\hbar}{8} \omega_{r}^{-1} (v_r/\Delta v)^{2}  (\Omega_{\text{arm1}})^2$.  The relevant quantity in determining the correction to the phase difference between the arms is the difference in energy shifts $\Delta E_{\text{arm1}} - \Delta E_{\text{arm2}}$, which is determined by the lattice depth imbalance between the arms.  Where we let $\Omega_{\text{arm1}} = \Omega$ and $\Omega_{\text{arm2}} = \Omega + \Delta \Omega$:

\begin{equation} \label{eqn:offresshift}
 \Delta E_{\text{arm1}} - \Delta E_{\text{arm2}} \sim \frac{\hbar}{4} \frac{\Omega \Delta \Omega}{\omega_{r} (\Delta v/v_r)^{2}}  \sim \hbar (0.4 ~\text{s}^{-1}) \left(\frac{2 \pi \times 3.77\text{kHz}}{\omega_{r}}\right)\left(\frac{20}{\Delta v/v_r} \right)^{2} \left(\frac{\Omega}{5 \omega_r} \right)^2 \left (\frac{\Delta \Omega/\Omega}{10^{-3}} \right)
 \end{equation}  
 
\begin{figure} \includegraphics[width=7.5in]{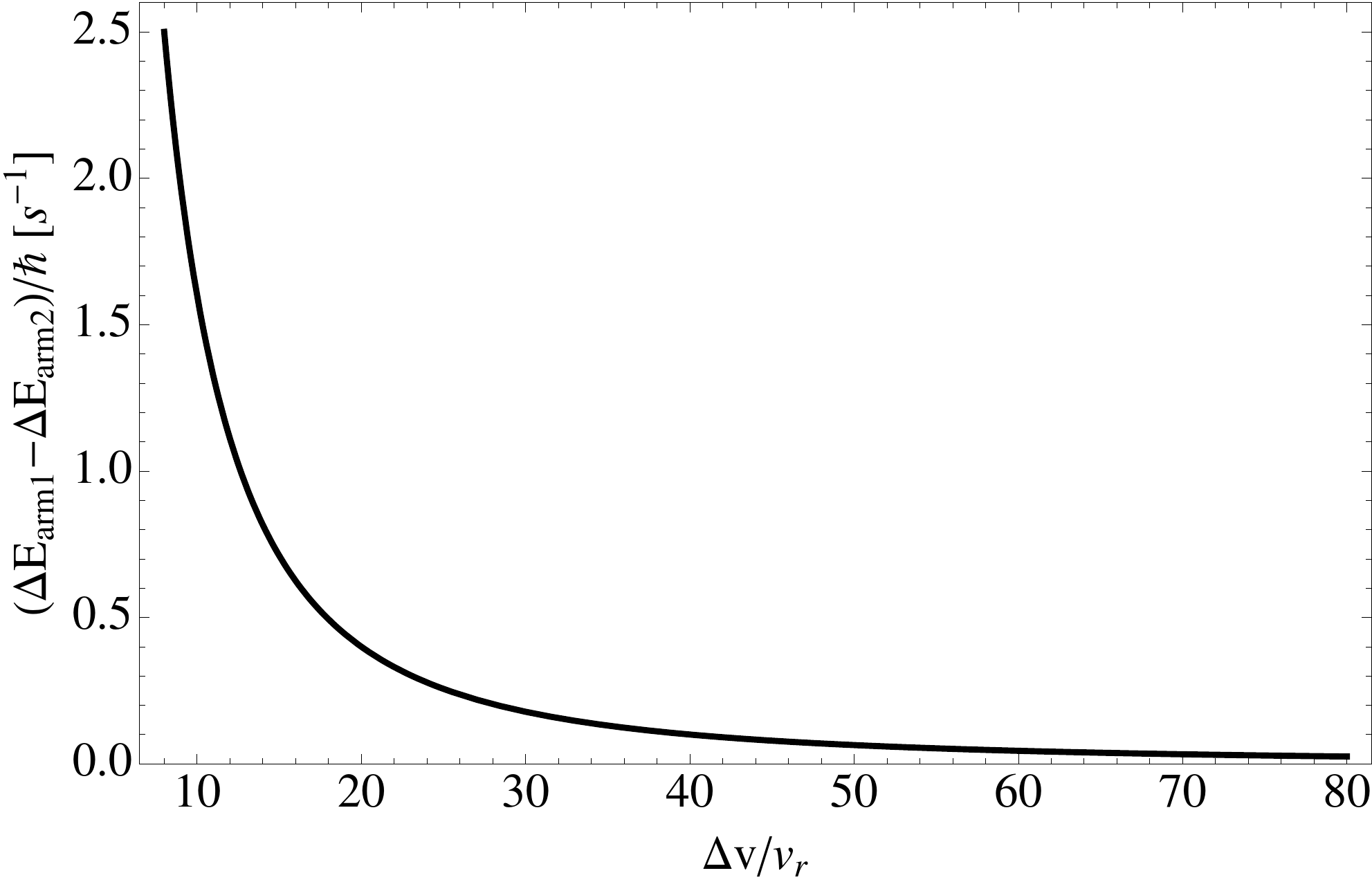} 
\caption{(color online) Dependence of $(\Delta E_{\text{arm1}} - \Delta E_{\text{arm2}})/\hbar$ on $(\Delta v/v_r)$ as given in Eq. \ref{eqn:offresshift}, which quantifies the effect of the off-resonant lattices.  This plot takes the lattice depth for arm 1 to be $\Omega_{\text{arm1}} = \Omega$ and the lattice depth for arm 2 to be  $\Omega_{\text{arm2}} = \Omega + \Delta \Omega$, where $\frac{\Delta \Omega}{\Omega}=10^{-3}$ and $\Omega = 5 \omega_r$.  The recoil frequency is taken to be $\omega_r = 2 \pi \times 3.77\text{kHz}$.}
\label{fig:shift}
\end{figure}

 \noindent This result agrees with numerical simulations.  Fig. \ref{fig:shift} illustrates the dependence of  $(\Delta E_{\text{arm1}} - \Delta E_{\text{arm2}})/\hbar$ on $(\Delta v/v_r)$.  For the same reasoning as in the discussion in the previous section, the net effect of lattice depth imbalances on the correction term treated here will often be further suppressed by orders of magnitude for interferometers in a differential configuration.  Thus, as stated previously, in many cases the net effect of phase corrections due to the off-resonant lattices will be below the mrad level in a differential configuration. 

\begin{figure}
\includegraphics[width=3.00in]{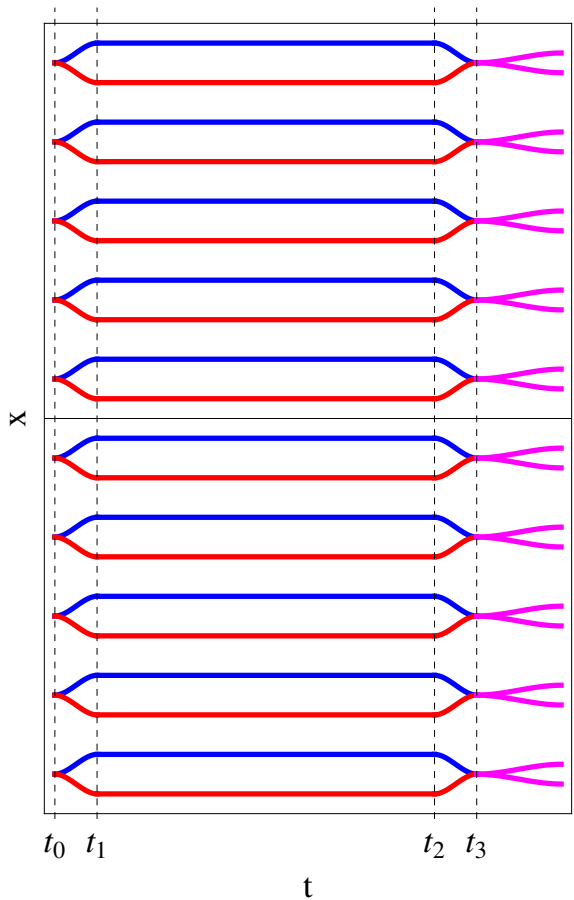}
\caption{(color online) An array of lattice gravimeters such as that shown above can be used to achieve measurements of a local gravitational field with high spatial resolution.  The trajectories shown hold in the lab frame.  After separating the atoms in each gravimeter by a small amount, we can implement a hold sequence to greatly increase sensitivity.  Bragg pulses are used at times $t_0$, $t_1$, $t_2$, and $t_3$.  Such an array could be used  to study general relativistic effects, search for extra dimensions, examine local mass distributions, or measure Newton's constant.  In addition, the ability of lattice interferometers with hold sequences to provide extremely precise measurements with a small interrogation region makes them ideal candidates for compact, mobile sensors.}
\label{fig:array}
\end{figure}

A lattice gravimeter can provide extraordinary levels of sensitivity.  This sensitivity can be achieved over small distance scales by implementing a hold sequence in which the two arms are separated, manipulated into the same momentum eigenstate, held in place by a single lattice, and then recombined.  In achieving compactness, we note that the fact that lattice interferometers are confined and can thus keep the atoms from falling under gravity during the separation and recombination stages of the interferometer as well as during the hold sequence is essential.  Otherwise, for many configurations, the desired arm separation could not be reached without the atoms falling too great a distance, which would ruin the compactness of the interferometer.  Hold times will be limited by spontaneous emission, which decreases contrast.  Modern laser technology will allow us to use detunings of hundreds or even thousands of GHz, making hold times on the order of 10 s within reach \cite{ref:gravitywaves}.  Gravimeter sensitivities using the hold method greatly exceed the sensitivities of light-pulse gravimeters while simultaneously allowing for a significantly smaller interrogation region.   For example, for $\sim 10^7$ atoms/shot and $\sim 10^{-1}$ shots/s, a shot noise limited conventional light-pulse interferometer with a 10 m interrogation region can achieve a sensitivity of $\sim 10^{-11}$ $g$/Hz$^{1/2}$.  With similar experimental parameters, a shot noise limited lattice interferometer with a 10 s hold time and an interrogation region of 1 cm will have a sensitivity of $ \sim 10^{-12}$ $g$/Hz$^{1/2}$.  If we expand the interrogation region to 1 m, we obtain a sensitivity of $\sim 10^{-14}$ $g$/Hz$^{1/2}$.  This remarkable sensitivity has a plethora of potential applications.  Extremely precise gravimeters and gravity gradiometers can be constructed to perform tests of general relativity, make measurements relevant to geophysical studies, and build highly compact inertial sensors.  Moreover, the fact that lattice interferometers can operate with such high sensitivities over small distance scales makes them prime candidates for exploring short distance gravity.  One could set up an array of lattice gravimeters to precisely map out gravitational fields over small spatial regions, as shown in Fig. \ref{fig:array}.  The knowledge obtained about the local gravitational field could be useful in searching for extra dimensions \cite{ref:nima} as well as in studying the composition and structure of materials.

\subsection{Tests of atom charge neutrality}

Ultra-high precision gravitational measurements are certainly among the most promising applications of lattice interferometers, but the usefulness of lattice interferometers is certainly not limited to the study of gravity.  By exposing the two arms of a lattice interferometer to different electrostatic potentials, tests of atom charge neutrality with unprecedented accuracy could be achieved \cite{ref:arvanitaki}.  The main advantage of a lattice interferometer in such a measurement is that the interrogation time can be significantly increased in comparison to the interrogation time achievable in a light-pulse geometry through the use of a hold sequence.  Where  $t_{\text{hold}}$ is the duration of the hold, $V$ is the electrostatic potential difference between the two arms of the interferometer, $e$ is the electron charge, and $\epsilon$ is the ratio of the atomic charge to the electron charge, the phase shift is given by $\frac{\epsilon e}{\hbar} V t_{\text{hold}}$ \cite{ref:arvanitaki}.  Based on the results of Sec. IV., Sec. V., and Appendix B, and assuming an identical configuration to that described in \cite{ref:arvanitaki} except for the inclusion of a hold sequence, we estimate that the phase error induced by the undesirable effects we consider will be below the proposed shot noise limit for this experiment (1 mrad).  Any systematic phase error can be characterized by the methods we have developed.  For a hold time of 10 s and for integration over $10^6$ shots, atom charges can be probed down to the region of $\epsilon \sim 10^{-27}$.

\subsection{Measurements of $\frac{\hbar}{m}$ and of isotope mass ratios}

\begin{figure}
\includegraphics[width=3.00in]{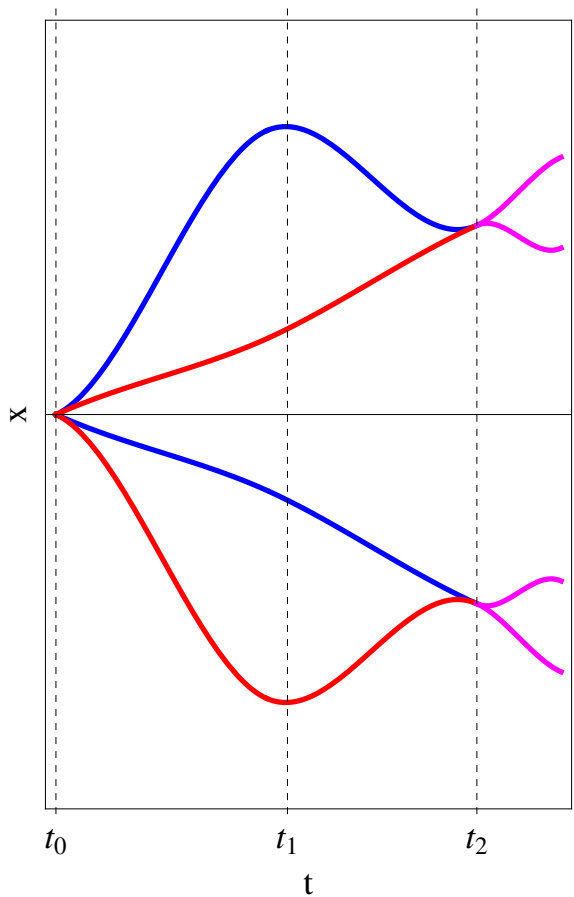}
\caption{(color online) Conjugate interferometer geometry that could be used to measure $\frac{\hbar}{m}$.  The trajectories shown hold in the lab frame.  Bragg pulses are used at times $t_0$, $t_1$, and $t_2$ (at $t_0$, a sequence of multiple Bragg pulses is necessary to split the system into the two arms of the two conjugate interferometers).  The phase shift of the lower interferometer is subtracted from the phase shift of the upper interferometer, which suppresses the effects of laser phase noise and eliminates the gravitational phase shift up to gradients \cite{ref:hermann}.  With such a scheme, a measurement of $\frac{\hbar}{m}$ to a part in $10^{14}$ may be possible, which could lead to the most accurate determination of the fine structure constant to date.}
\label{fig:conj}
\end{figure}

The ratio $\frac{\hbar}{m}$ is of particular interest because of its direct relation to the fine structure constant.  Atom interferometry has previously been used to provide exquisite measurements of $\frac{\hbar}{m}$.  The most precise atom interferometric measurement to date was performed by Cadoret {\it et al.}, who performed an elegant experiment combining Bloch oscillations and a Ramsey-Bord\'{e} interferometer to measure the fine structure constant  to within a relative uncertainty of $4.6 \times 10^{-9}$ \cite{ref:cadoret}.  Eq. (\ref{eqn:phaseshift}) indicates that if we apply different accelerations to the two arms of the interferometer, we will see a phase shift proportional to $\frac{m}{\hbar}$ that  depends on the kinetic energy difference between the arms, which can be made extremely large.  A differential configuration using conjugate interferometers (shown in Fig. \ref{fig:conj}) could reduce the net contribution of such unwanted effects as laser phase noise and cancel the gravitational phase shift up to gradients \cite{ref:hermann}.  Such a geometry could provide an extremely precise measurement of $\frac{\hbar}{m}$, as illustrated by the fact that we can achieve a phase shift of $\sim 10^{11}$ radians for a 5 m interrogation region and a 0.6 s interrogation time, corresponding to a shot noise limited sensitivity of $\sim 10^{-14} \frac{\hbar}{m}$/Hz$^{1/2}$ for the experimental parameters stated above.  In this situation, the dominant unwanted effect would arise from non-adiabatic corrections, and the methods for calculating these corrections that are presented in the previous sections would need to be applied (we estimate a phase error $\sim$ 10 rad).  We note that even if the shot noise limit is not reached, the technique we have proposed could still improve over current $\frac{\hbar}{m}$ measurements.  We emphasize again that to take full advantage of the sensitivity offered by lattice manipulations in atom interferometry, the methods we develop in this paper for calculating non-adiabatic corrections are absolutely essential.

The fact that all terms in Eq. (\ref{eqn:phaseshift}) are proportional to $m$ (except for the $\phi_{\text{Bragg}}$ term, which as we have explained, will often be negligible) can be exploited to provide high-precision measurements of isotope mass ratios by using an interferometer geometry in which the two isotopes follow identical trajectories.  Isotope mass ratios could be relevant to studies of advanced models of the structure of the nucleus \cite{ref:lunney}.  Conversely, if we know the mass ratio of two isotopes sufficiently well, we can use such a geometry to precisely measure accelerations, where the two isotopes provide two dissimilar conjugate interferometers.  The phase noise of these two interferometers will be extremely well correlated because they are topologically identical.  This also eliminates the need for the additional lattice beams required to form the second, topologically distinct interferometer that would be needed if we only had a single isotope.  Note that this scheme to construct dissimilar, topologically identical conjugate accelerometers would not be possible for a light-pulse geometry, for the leading order phase shift of light-pulse accelerometers is independent of isotope mass.

\subsection{Gyroscopes}

Lattice interferometers can also be used to build compact and highly sensitive gyroscopes.  There are multiple possible schemes in which optical lattices can enhance gyroscope sensitivity.  One such scheme is to modify a typical atom-based gyroscope by replacing the Raman pulses with LMT lattice beam splitters, increasing the enclosed area of the interferometer and hence its sensitivity to rotations.  The phase shift can be written in Sagnac form as $\frac{2m}{\hbar} \vec{\Omega} \cdot \vec{A}$, where $\vec{\Omega}$ is the rotation rate vector and $\vec{A}$ is the normal vector corresponding to the enclosed area of the interferometer \cite{ref:varenna}.  The gyroscope described in \cite{ref:gustavson} achieves a sensitivity of $2 \times 10^{-8} \frac{\text{rad}}{\text{s}}$/Hz$^{1/2}$.  Replacing the Raman pulses in this experiment with $200 \hbar k$ lattice beam splitters would increase the sensitivity by a factor of 100.  In such a configuration, we estimate that each beam splitter could introduce a non-adiabatic phase error of $\sim$ 1 rad if the arms of the interferometer are not split symmetrically.  However, beam splitter configurations that exploit symmetry between the arms can reduce this effect by orders of magnitude.  Another option is to use optical lattices along multiple axes to provide complete control of the motion of the atoms in two or three dimensions (this control is only achieved in the region in which the lattices overlap, necessitating the use of wide beams).  Analagous to a fiber-optic gyroscope, the atoms could be guided in repeated loop patterns, with the two arms rotating in opposite directions.  Geometries in which atomic motion is controlled in multiple dimensions could also expand the possibilities for other applications of lattice interferometry (such as measurements of gravity) by allowing for the measurement of potential energy differences between arbitrary paths.  For instance, a compact array of three orthogonal lattice gravity gradiometers could be used to measure the nonzero divergence of the gravitational field in free space predicted by general relativity \cite{ref:gr}.

\section{Conclusion}

We have presented a detailed analytical description of the interaction between an atom and an optical lattice, using the adiabatic approximation as a starting point and then proceeding to rigorously develop a method to calculate arbitrarily small corrections to this approximation using perturbative adiabatic expansion.  We have applied this theoretical framework to calculate the phase accumulated during a lattice acceleration in an LMT beam splitter.  And we have proposed atom interferometer geometries that use optical lattices as waveguides and discussed applications of such geometries, using our theoretical methods to add rigor to this discussion.  We are working toward the experimental implementation of lattice interferometers and LMT lattice beam splitters, and we hope to explore the applications we have discussed.  In this experimental work, we realize that we will have to contend with a number of unwanted systematic effects, such as spatially varying magnetic fields, imperfections in the lattice beam wavefronts, and inhomogeneity of the lattice depth across the atomic cloud.  We have studied these effects using both analytical and numerical methods, and we are optimistic that they can be significantly mitigated for a wide range of experimental parameters--a conclusion that we hope to verify experimentally.  Many of the unwanted systematic effects that are relevant to lattice interferometers are also shared by light-pulse interferometers and can therefore be dealt with using similar methods.  Therefore, we believe that it will likely be possible to realize lattice interferometers in existing apparatuses originally constructed with light-pulse geometries in mind.

\section*{ACKNOWLEDGMENTS} 

We would like to thank Philippe Bouyer, Sheng-wey Chiow, Gerald Gabrielse, and Holger Mueller for valuable discussions.  TK acknowledges support from a Fannie and John Hertz Foundation Fellowship and an NSF Fellowship.  TK and DJ acknowledge support from a Stanford Graduate Fellowship. 

\appendix

\section{BOOSTING BETWEEN DIFFERENT FRAMES}

For the purposes of this paper, we consider unitary transformations that consist of a translation in position space, a boost in momentum space, and a time-dependent change of phase.  Such a transformation has the general form:

\begin{equation}
\ \hat{U}(t) = e^{\frac{i}{\hbar} d(t) \hat{p}} e^{- \frac{i}{\hbar} m v(t) \hat{x}} e^{\frac{i}{\hbar} \theta(t)}
\end{equation}

\noindent We note we are free to choose the boost parameters $d(t)$ and $v(t)$ arbitrarily.  That is, we do not have to choose them so that $v(t)$ is the rate of change of $d(t)$.  The operators $\hat{x}$ and $\hat{p}$ transform under $\hat{U}(t)$ as follows:

\begin{eqnarray}
\nonumber \hat{U}(t) \hat{x} \hat{U}^{\dag}(t) &=&\hat{x} + d(t) \\
\nonumber \\
\nonumber \hat{U}(t) \hat{p} \hat{U}^{\dag}(t) &=&\hat{p} +mv(t) \\
\end{eqnarray}

\noindent We now consider the Hamiltonian in a frame that is freely falling with gravity, which takes the form:

\begin{equation}
\ \hat{H}_{\text{FF}} =\frac{\hat{p}^2}{2m} + 2 \hbar \Omega(t) \sin^2 \left [k\hat{x}-k (D_{\text{Lab}}(t)+ \frac{1}{2} g t^2)\right ]
\end{equation}

\noindent We can transform from the freely falling frame to the lab frame by applying the appropriate Galilean transformation $\hat{U}_{\text{FF}}(t)$, which corresponds to specifying $d(t) = \frac{1}{2} g t^2$, $v(t) = gt$, and $\theta(t) = \frac {1}{3} m g^2 t^3$, so that $\hat{H}_{Lab} = \hat{U}_{FF}(t) \hat{H}_{FF} \hat{U}^\dag_{FF}(t) + i \hbar \left ( \frac{\partial}{\partial t} \hat{U}_{FF}(t) \right) \hat{U}^\dag_{FF}(t)$.  

Since the only position dependence in $\hat{H}_{\text{FF}}$ comes from the lattice potential, we could readily approach the problem of finding the dynamics of the system by working in the freely falling frame.  However, it will prove to be useful from a calculational standpoint to transform to a third frame, with a Hamiltonian resembling that describing the atom-light interaction from the point of view of dressed states.  We note that although we could have performed a boost directly from the lab frame to the dressed state frame, it is useful to introduce the freely falling frame for pedagogical reasons, since it is the frame in which calculations for atom interferometry are typically performed.  In Appendix B, we use the transformation between the freely falling frame and the dressed state frame to highlight the parallels between a lattice beam splitter and a typical light pulse beam splitter.  We absorb the initial velocity $v_0$ of the atom in the lab frame (and hence also in the freely falling frame) into the dressed state frame, so that velocity $v_0$ in the lab frame corresponds to velocity zero in the dressed state frame.  The Hamiltonian in the dressed state frame is:

\begin{equation}
\ \hat{H}_{\text{DS}} =\frac{\hat{p}^2}{2m} - (v_{\text{Lab}}(t) + gt - v_0) \hat{p} + 2 \hbar \Omega(t) \sin^2 \left(k\hat{x} \right)
\end{equation}

\noindent The unitary transformation $\hat{U}_{\text{DS}}$ that transforms from the dressed state frame to the freely falling frame, so that $\hat{H}_{FF} = \hat{U}_{DS}(t) \hat{H}_{DS} \hat{U}^\dag_{DS}(t) + i \hbar \left ( \frac{\partial}{\partial t} \hat{U}_{DS}(t) \right) \hat{U}^\dag_{DS}(t)$, corresponds to boost parameters $d(t) = -(D_{\text{Lab}}(t) + \frac{1}{2} g t^2)$ and $v(t) = - v_0$ with a time-dependent phase factor:

\begin{equation} \label{eqn:dstransform}
\ \hat{U}_{\text{DS}} =e^{\frac{i}{\hbar} mv_0 \hat{x}} e^{- \frac{i}{\hbar} (D_{\text{Lab}}(t) + \frac{1}{2} g t^2) \hat{p}} e^{- \frac{i}{\hbar} (\frac{1}{2} m v_0^2 t)}
\end{equation}

\noindent  The ability to boost to a frame in which the Hamiltonian contains no position dependent terms outside of the lattice potential is contingent upon the assumption that the external potential in the lab frame (not including the lattice potential) is linear in $x$.  However, real-world potentials such as the potential corresponding to Earth's gravitational field will deviate somewhat from this assumption.  Any such deviations would manifest as residual position dependent terms in the dressed state Hamiltonian, which we collectively refer to as $V^{\prime}$.  Under the semiclassical approximation, we neglect the effects of $V^{\prime}$ on the time evolution of the atomic wavepacket.  This approximation is valid when the energy scale of $V^{\prime}$ over the spread of the atom's wavefunction (which is on the order of magnitude of the expectation value of $V^{\prime}$ in the atomic wavepacket) is much smaller than the energy scale of the lattice potential and is small relative to the time scale of the experiment, which is the case for a wide class of experimental parameters.  For example, in the case of a Rubidium atom wavepacket with a spatial spread of $\sigma \sim 100 \mu$m in the gravitational field at the Earth's surface (which has a gradient of $\lambda \sim10^{-6} \; s^{-2}$), the energy scale of $V^{\prime}$ will be $\sim \frac{1}{2} m \lambda \sigma^2 \sim h$ $\times$ ($10^{-6}$ Hz).  This energy scale is smaller than that of a lattice of typical experimental depth ($\sim$ $5 E_{r}$, where $E_{r}$ is the recoil energy $\frac{\hbar^2 k^2}{2m})$ by a factor of $\sim 10^{10}$ and is small on a time scale of $\sim 10$ s.  The effects of linear gradients and of more general potentials can be accounted for through a straightforward generalization of the results presented in this paper, as discussed in greater depth in \cite{ref:kovachy}.

\section{GENERALIZING TO THE CASE OF A FINITE WAVEPACKET}

In Sec. II., we discretized the Hamiltonian using the basis of momentum states $\left | 2n\hbar k \right >$ for integer $n$.  We now generalize our results to the case of a finite wavepacket.

Throughout our analysis, we have worked mainly in the dressed state frame, since this frame is particularly convenient for describing phase evolution in a lattice.  For interferometers where we use lattices as waveguides for the atoms, we will want to perform the entire phase shift calculation in this frame.  However, phase shift calculations for light-pulse atom interferometers are often performed in the freely falling frame.  Thus, for applications where lattice manipulations are used for beam splitters and mirrors in a light-pulse geometry, it is useful to convert the phase evolution we calculate in the dressed state frame to the freely falling frame.  We thus present the general results derived in this appendix in the freely falling frame.

In Sec. II., we considered a particular dressed frame in which the initial velocity of the atom is boosted to zero.  We now introduce an unboosted dressed state frame that is related to the freely falling frame by a translation in position space with no boost in momentum space, so that the unitary transformation $\hat{U}_{\text{DS}_0} =e^{- \frac{i}{\hbar} (D_{\text{Lab}}(t) + \frac{1}{2} g t^2) \hat{p}}$ transforms from the unboosted dressed state frame to the freely falling frame.  The Hamiltonian in this frame is:

\begin{equation}
\ \hat{H}_{\text{DS}_0} =\frac{\hat{p}^2}{2m} - (v_{\text{Lab}}(t) + gt) \hat{p} + 2 \hbar \Omega(t) \sin^2 \left(k\hat{x} \right)
\end{equation}

Now, say that before the lattice acceleration, the state that we are accelerating is described by $\left | \Psi_{\text{FF}}(t_0) \right >$ in the freely falling frame.  We can then transform this state vector to the unboosted dressed state frame, describe its evolution to the final time $t_f$ in this frame, and transform back to the freely falling frame.  Where $\hat{T}_{\text{DS}_0}(t^{\prime},t)$ is the time evolution operator the takes us from time $t$ to time $t^{\prime}$ in the unboosted dressed state frame, we can write:

\begin{equation} \label{eqn:evolution}
\left | \Psi_{\text{FF}}(t_f) \right > =\hat{U}_{\text{DS}_0}(t_f) \hat{T}_{\text{DS}_0}(t_f,t_0)  \hat{U}^{\dag}_{\text{DS}_0}(t_0) \left | \Psi_{\text{FF}}(t_0) \right >
\end{equation}

\noindent Denoting the initial momentum space wavefunction in the freely falling frame as $\vec{\Psi}_{\text{FF}}(p,t_0) \equiv \left < p | \Psi_{\text{FF}}(t_0) \right >$, we can express Eq. (\ref{eqn:evolution}) as:

\begin{eqnarray} \label{eqn:freelyfallingevolution}
\nonumber \left | \Psi_{\text{FF}}(t_f) \right > &=& \hat{U}_{\text{DS}_0}(t_f) \hat{T}_{\text{DS}_0}(t_f,t_0)  \hat{U}^{\dag}_{\text{DS}_0}(t_0) \int dp\left | p \right > \vec{\Psi}_{\text{FF}}(p,t_0) \\
\nonumber &=& \int dp \hat{U}_{\text{DS}_0}(t_f) \hat{T}_{\text{DS}_0}(t_f,t_0) \left | p \right > e^{ \frac{i}{\hbar} (D_{\text{Lab}}(t_0) + \frac{1}{2} g t_0^2) p} \vec{\Psi}_{\text{FF}}(p,t_0) \\
\end{eqnarray}

\noindent  where we have used the linearity of the operators to bring them inside the integral.  We now consider how each momentum eigenstate $\left | p \right >$ evolves in the unboosted dressed state frame.    That is, we must calculate $\hat{T}_{\text{DS}_0}(t_f,t_0) \left | p \right >$ for each $p$.  In order to do so, we introduce a class of boosted dressed state frames $DS_p$ parameterized by $p$, so that momentum $p$ in the unboosted dressed state frame (which is just the frame $DS_0$) corresponds to momentum zero in the frame $DS_p$.  In essence, where $v \equiv \frac{p}{m}$, frame $DS_p$ travels with velocity $v$ with respect to the unboosted dressed state frame.  In frame $DS_p$, the Hamiltonian takes the form:

\begin{equation}
\hat{H}_{\text{DS}_p} =\frac{\hat{p}^2}{2m} - (v_{\text{Lab}}(t) + gt-v) \hat{p} + 2 \hbar \Omega(t) \sin^2 \left(k\hat{x} \right)
\end{equation}

\noindent  where we note that the unitary transformation that transforms from frame $DS_p$ to the unboosted dressed state frame is:

\begin{equation}
\hat{U}_p(t) =e^{\frac{i}{\hbar} p \hat{x}}e^{- \frac{i}{\hbar}\left[-(D_{\text{Lab}}(t) + \frac{1}{2} g t^2) mv + \frac{1}{2}mv^2 t\right]}
\end{equation}

\noindent  Where $\hat{T}_{\text{DS}_p}(t_f,t_0)$ is the time evolution operator in frame $DS_p$, we can write:

\begin{eqnarray} \label{eqn:boostedds}
\nonumber \hat{T}_{\text{DS}_0}(t_f,t_0) \left | p \right > &=& \hat{U}_p(t_f) \hat{T}_{\text{DS}_p}(t_f,t_0) \hat{U}^{\dag}_p(t_0) \left | p \right > \\
\nonumber  &=& e^{\frac{i}{\hbar}\left[-(D_{\text{Lab}}(t_0) + \frac{1}{2} g t_0^2) mv + \frac{1}{2}mv^2 t_0\right]} \hat{U}_p(t_f) \hat{T}_{\text{DS}_p}(t_f,t_0) \left | 0 \right > \\
\end{eqnarray}

The problem of determining the evolution of the momentum eigenstate $\left | p \right >$ in the unboosted dressed state frame thus reduces to evolving the momentum eigenstate $\left | 0 \right >$ in the frame $DS_p$, which we know how to do from the preceding sections.  We assume that the momentum space wavefunction is narrow enough so that the momentum eigenstates we consider are resonant with the lattice (that is, as we recall from Sec. III., the magnitudes of their velocities with respect to the lattice are less than $v_r$).  Where the lattice acceleration is chosen so as to transfer a momentum of $2n \hbar k$ to the atom, the result from Sec. III. tells us that the time evolution in Eq. (\ref{eqn:boostedds}) yields:

\begin{equation}
\hat{T}_{\text{DS}_p}(t_f,t_0) \left | 0 \right > = e^{i \phi_p} \left | 2n \hbar k \right >
\end{equation}

\noindent for phase:

\begin{equation} \label{eqn:velocityphase}
\phi_p = \frac{m}{2 \hbar}\int_{t_0}^{t_f}v_{\text{Lattice}}^p(t)^2d t
\end{equation}

\noindent where $v_{\text{Lattice}}^p(t) \equiv v_{\text{Lab}}(t) + gt-v$ is the lattice velocity in frame $DS_p$.  For the sake of pedagogy, at the moment we neglect the phase arising from the lattice depth, the phase arising from the small periodic variations in the lowest eigenvalue, and the phase arising from non-adiabatic corrections, where we note that we could calculate these contributions if needed.  Evaluating the integral in Eq. (\ref{eqn:velocityphase}), we obtain:

\begin{equation} \label{eqn:boostedphase}
\phi_p = \frac{m}{2 \hbar}\int_{t_0}^{t_f}(v_{\text{Lab}}(t) + gt)^2d t-\frac{m}{\hbar}\left[(D_{\text{Lab}}(t_f) + \frac{1}{2}gt_f^2)-(D_{\text{Lab}}(t_0) + \frac{1}{2}gt_0^2)\right]v+ \frac{m}{2 \hbar}v^2(t_f-t_0)
\end{equation}

\noindent Substituting this result into Eq. (\ref{eqn:boostedds}), applying the transformation $ \hat{U}_p(t_f)$, and canceling terms in the exponentials, we find that:

\begin{equation} \label{eqn:ds0phase}
\hat{T}_{\text{DS}_0}(t_f,t_0) \left | p \right > = e^{i \frac{m}{2 \hbar}\int_{t_0}^{t_f}(v_{\text{Lab}}(t) + gt)^2d t} \left | p  + 2n\hbar k\right > \end{equation}

\noindent We can now evaluate Eq. (\ref{eqn:freelyfallingevolution}), which gives us the final result:

\begin{equation}
\left | \Psi_{\text{FF}}(t_f) \right > = \int dp\left | p +2n \hbar k \right > e^{i\phi_{\text{FF}}(p)} \vec{\Psi}_{\text{FF}}(p,t_0)
\end{equation}

\noindent where:

\begin{equation} \label{eqn:pulsephase}
\phi_{\text{FF}}(p)=\frac{m}{2 \hbar}\int_{t_0}^{t_f}(v_{\text{Lab}}(t) + gt)^2dt - 2nk (D_{\text{Lab}}(t_f) + \frac{1}{2} g t_f^2) -\frac{p}{\hbar} \Delta D
\end{equation}

\noindent for $\Delta D \equiv (D_{\text{Lab}}(t_f) + \frac{1}{2} g t_f^2) - (D_{\text{Lab}}(t_0) + \frac{1}{2} g t_0^2)$.

Observe that the second term in Eq. (\ref{eqn:pulsephase}) is what would typically be called the laser phase in a light-pulse atom interferometer for an $n$-photon beam splitter.  We note that as long as the resonance condition is met for the momentum eigenstates we are considering so that they are accelerated, the phase evolved during a lattice acceleration in the $\text{DS}_0$ dressed state frame, is independent of $p$, as shown in Eq. (\ref{eqn:ds0phase}).  (This is true up to non-adiabatic corrections and the small periodic variations in the ground state eigenvalue).  This makes dressed state frames particularly convenient for performing calculations involving wavepackets.  In contrast, in the freely falling frame, the accumulated phase $\phi_{\text{FF}}$ is dependent on $p$, but this dependence cancels in the final expression for the phase shift between the two arms of an interferometer as long as the distance travelled by the atom while locked into a lattice is the same for both arms.  Note that momentum dependent contributions to the total phase shift must arise when we treat the problem purely in terms of dressed states, since the total phase shift is an observable quantity and must therefore be independent of the frame in which it is calculated.  The key point to realize is that if the total distance traveled in the lattice is not the same for both arms, then the two arms will end up in two different dressed state frames.   

We now consider the effect on phase evolution of the periodic term $p_0(\alpha,\tilde{\Omega})$ in the expression for the ground state eigenvalue of the dressed state Hamiltonian given in Eq. (\ref{eqn:evalue}).  This term leads to a momentum dependent correction $\delta \phi_p$ to the evolved phase $\phi_p$ described in Eq. (\ref{eqn:velocityphase}), where:

\begin{equation} \label{eqn:periodicphase}
\delta \phi_p = -\frac{1}{\hbar}\int_{t_0}^{t_f} p_0\left(\frac{v_{\text{Lattice}}^p(t)}{v_r},\tilde{\Omega}(t) \right) d t
\end{equation}

\noindent We note that $p_0\left(\frac{v_{\text{Lattice}}^p(t)}{v_r},\tilde{\Omega}(t) \right)$ can be calculated using the truncated matrix approximation described in Sec. IV.  Since $p_0\left(\frac{v_{\text{Lattice}}^p(t)}{v_r},\tilde{\Omega}(t) \right)$ is periodic in $v_{\text{Lattice}}^p(t)$ with period $2 v_r$, the contribution $\delta \phi_p^{\text{arm1}} - \delta \phi_p^{\text{arm2}}$ of this correction term to the phase difference between the two arms of an interferometer will be highly suppressed if both arms load the atom into the lattice near the center of the zeroth band (as discussed in Sec. III.) and undergo a nearly integral number of Bloch oscillations so that the effects of this periodic variation in the ground state eigenvalue will be largely common to the two arms, as verified by estimation of the integral in Eq. (\ref{eqn:periodicphase}) and by numerically solving the Schrodinger equation.

\section{THE RESONANCE CONDITION FOR AN ATOM TO BE LOADED INTO THE GROUND STATE OF A LATTICE}

Here, we derive the resonance condition stated in Sec. III., which says that for a sufficiently slow lattice depth ramp, the atom will be loaded into the ground state of the lattice if the initial velocity of the lattice is within $v_r$ of the initial velocity of the atom.  First, we examine the eigenvalues of $H$ when $\tilde{\Omega}(t) = 0$.  For  $\tilde{\Omega}(t) = 0$, $H$ is diagonal matrix.  Since $H$ is a matrix defined in terms of the basis of momentum eigenstates $\left | 2n\hbar k \right >$ for integer $n$, when $H$ is diagonal these momentum eigenstates are also the eigenstates of $H$, with corresponding eigenvalues $E_n = 4 E_r \left (n^2 - n \alpha(t) \right )$ (which are just the diagonal elements of $H$).  We note that the effect of the $- (v_{\text{Lab}}(t) + gt - v_0) \hat{p}$ term in the Hamiltonian is to encapsulate the dependence of the eigenvalues and eigenvectors of the Hamiltonian on the lattice velocity for any given lattice depth, which can be best conceptually understood from the point of view of the dressed state picture explained in \cite{ref:peik}.  The resonance condition states that if the initial state is $\left | 2n_0 \hbar k \right >$, then the initial velocity of the lattice must be within $v_r$ of $2 n_0 v_r$.  Thus, the initial value $\alpha_0$ of the dimensionless velocity $\alpha$ must be in the range $(2n_0-1,2n_0+1)$.  We will now show why this condition on  $\alpha_0$ implies that the ground state of $H$ with $\tilde{\Omega}(t) = 0$ is $\left | 2n_0 \hbar k \right >$.  Where we write $\alpha_0 = 2 n_0 + \delta$ and let $n = n_0 +\Delta n$ for integer $\Delta n$, the eigenvalues of $H$ with the lattice depth set to zero are:

\begin{equation}
\ E_n = 4 E_r \left[ \left(n_0 + \Delta n \right) ^2 - \left (n_0 + \Delta n \right) \left (2 n_0 +\delta \right) \right]= 4 E_r \left[ -n_0^2 + (\Delta n)^2 -n_0 \delta- \Delta n \delta \right]
\end{equation}

\noindent For $| \delta | < 1$, $E_n$ will be most negative when $\Delta n = 0$, so $\left | 2n_0 \hbar k \right >$ will indeed be the ground state of the Hamiltonian when $\tilde{\Omega}(t) = 0$.  Therefore, as long as we ramp up the lattice slowly enough and keep the lattice velocity constant while doing so, the adiabatic theorem tells us that the atom will remain in the lattice ground state throughout the ramp up process provided that the ground state never passes through a point of degeneracy, which is indeed the case for fixed $\alpha$ and $| \delta | < 1$.

The maximum rate at which we can increase the lattice depth while still maintaining the validity of the adiabatic approximation depends on $| \delta |$.  This statement follows from the adiabatic condition $\left | \left < 1 \right | \dot{H} \left | 0 \right > \right | \ll \frac{(\varepsilon_1-\varepsilon_0)^2}{\hbar}$, where $\left | 0 \right >$ and $\left | 1 \right >$ respectively denote the ground state and the first excited state of the Hamiltonian.  When the resonance condition holds, note that the first excited state will be $\left | 2(n_0+\text{sign}(\delta)) \hbar k \right >$.  As $| \delta |$ increases toward 1, the gap between $E_{n_0}$ and $E_{n_0+\text{sign}(\delta)}$ (i.e. the energy gap between the zeroth and first bands) decreases.  To see why this is true, we observe that:

\begin{equation}
\ E_{n_0+\text{sign}(\delta)} - E_{n_0} = 4 E_r (1 - \text{sign}(\delta)  \delta) = 4 E_r (1 - |\delta |)
\end{equation}

\noindent  Therefore, for the adiabatic approximation to hold, the maximum rate at which we can ramp up the lattice decreases as we increase $| \delta |$.  This result is identical to the statement in Sec. III. that adiabatic loading is more difficult to achieve near the border of the first Brillouin zone.

After ramping up the lattice, we accelerate it until its velocity is within $v_r$ of $2 n_f v_r$ where $\left | 2n_f \hbar k \right >$ is the target momentum eigenstate.  The ground state of the Hamiltonian when the lattice has finished ramping down will therefore be $\left | 2n_f \hbar k \right >$, and as long as the entire process is adiabatic so that the atom remains in the ground state, the atom will indeed end up in the target state.  By analogy to the previous discussion, the maximum rate at which we can adiabatically ramp down the lattice increases as the difference between the final lattice velocity and $2 n_f v_r$ goes to zero.

\section{PERTURBATIVE ADIABATIC EXPANSION AT HIGHER ORDERS}

In order to calculate non-adiabatic corrections at arbitrary order, we define $C_{a \rightarrow b \rightarrow \cdots \rightarrow m \rightarrow n \rightarrow j}$ recursively in the natural way based on the notation of Sec. IV:

\begin{equation} \label{eqn:correctionarbitrary}
C_{a \rightarrow b \rightarrow \cdots \rightarrow m \rightarrow n \rightarrow j} \equiv -\int_{t_0}^{t} C_{a \rightarrow b \rightarrow \cdots \rightarrow m \rightarrow n} \vec{\Psi}^{\dag}_j(t^{\prime}) \frac{\partial \vec{\Psi}_n(t^{\prime})}{\partial t^{\prime}} e^{i \left[\varphi_n(t^{\prime}) - \varphi_j(t^{\prime}) \right ]} dt^{\prime}
\end{equation}

\noindent  For arbitrary order $k$, we can then write:

\begin{equation} \label{eqn:sumexpansion}
b^{(k)}_j(t) = C_j + \sum_{n \neq j} C_{n \rightarrow j} +  \sum_{n \neq j} \sum_{m \neq n} C_{m \rightarrow n \rightarrow j} + \; \cdots \; + \sum_{n \neq j} \sum_{m \neq n} \cdots \sum_{a \neq b}  C_{a \rightarrow b \rightarrow \cdots \rightarrow m \rightarrow n \rightarrow j}
\end{equation}

\noindent  where the last term includes $k$ sums and $ C_j \equiv b^{(0)}_j$.  In order for the expansion to be practical from a calculational standpoint, the series must converge quickly enough so that we do not have to find an unreasonable amount of terms.  There are two types of terms that we can often neglect.  First, where we assume that the system is initially in the ground state, it is often possible to neglect many terms of the form $C_{0 \rightarrow n}$ so that we only need to examine a relatively small number of eigenvectors.  Terms of this form indeed decrease rapidly as $n$ increases because the overlap of the $n$th eigenstate with the zeroth eigenstate is essentially nonexistent for large enough $n$.  Second, we can often neglect all terms of order greater than some cutoff value.  We will be able to do so as long as the time scale over which we are solving the problem is small enough so that integrating a $k$th order correction term against a factor of the form $\vec{\Psi}^{\dag}_j(t) \frac{\partial \vec{\Psi}_n(t)}{\partial t}$ yields a $(k+1)$th order correction term that is much smaller than the correction term of order $k$.

Note an important nuance in how we have phrased the above condition--we have mentioned nothing about a Hamiltonian that varies slowly in time.  The convergence of the perturbative series depends on the time interval of the solution.  As long as the Hamiltonian does not vary at an infinitely fast rate, we can always work on a small enough time scale so that the series converges rapidly.  To solve the problem on time scales for which the series does not converge quickly, we can simply break the problem into multiple parts.  This method provides with us with a means to describe the system for a Hamiltonian that changes arbitrarily fast in time.  Having a slowly varying Hamiltonian just serves to allow us to solve the problem without dividing it into as many parts (much of the time we will not have to divide the problem at all, and in Sec. V. we derive conditions for when this will be the case), thus making the calculation significantly easier.

\end{document}